\documentclass{aa}

\usepackage[utf8]{inputenc}
\usepackage[varg]{txfonts}
\usepackage{graphicx}
\usepackage{units}
\usepackage{siunitx}
\usepackage{textgreek}

\begin{document}

\title{Precise radial velocities of giant stars}

\subtitle{XV. Mysterious nearly periodic radial velocity variations in the eccentric binary {\bf \textepsilon}\,Cygni\thanks{Based on observations collected at the Lick Observatory, University of California.},\thanks{Based on observations collected with the Hertzsprung SONG telescope at the Spanish Observatorio del Teide, Tenerife.},\thanks{Based on data collected by the BRITE Constellation satellite mission, designed, built, launched, operated, and supported by the Austrian Research Promotion Agency (FFG), the University of Vienna, the Technical University of Graz, the University of Innsbruck, the Canadian Space Agency (CSA), the University of Toronto Institute for Aerospace Studies (UTIAS), the Foundation for Polish Science \& Technology (FNiTP MNiSW), and National Science Centre (NCN).},\thanks{RV data (Tables A.1 and A.2) are only available in electronic form at the CDS via anonymous ftp to \url{cdsarc.u-strasbg.fr} (\url{130.79.128.5}) or via \url{http://cdsweb.u-strasbg.fr/cgi-bin/qcat?J/A+A/}.}}

\author{Paul Heeren\inst{1}\thanks{\email{pheeren@lsw.uni-heidelberg.de}}\fnmsep\thanks{Fellow of the International Max Planck Research School for Astronomy and Cosmic Physics at the University of Heidelberg (IMPRS-HD)}
\and
Sabine Reffert\inst{1}
\and
Trifon Trifonov\inst{2}
\and
Ka Ho Wong\inst{3}
\and
Man Hoi Lee\inst{3,4}
\and
Jorge Lillo-Box\inst{6}
\and
Andreas Quirrenbach\inst{1}
\and
Torben Arentoft\inst{5}
\and
Simon Albrecht\inst{5}
\and
Frank Grundahl\inst{5}
\and
Mads Fredslund Andersen\inst{5}
\and
Victoria Antoci\inst{7,5}
\and Pere L. Pall\'e\inst{8,9}
}

\institute{Landessternwarte, Zentrum f\"ur Astronomie der Universit\"at Heidelberg, K\"onigstuhl 12, 69117 Heidelberg, Germany
\and
Max-Planck-Institut f\"ur Astronomie, K\"onigstuhl 17, 69117 Heidelberg, Germany
\and
Department of Earth Sciences, The University of Hong Kong, Pokfulam Road, 40000 Hong Kong, PR China
\and
Department of Physics, The University of Hong Kong, Pokfulam Road, Hong Kong, PR China
\and
Stellar Astrophysics Centre, Department of Physics and Astronomy, Aarhus University, Ny Munkegade 120, DK-8000 Aarhus C, Denmark
\and
Centro de Astrobiolog\'ia (CAB, CSIC-INTA), Depto. de Astrof\'isica, ESAC campus 28692 Villanueva de la Ca\~nada (Madrid), Spain\label{cab}
\and
DTU Space, National Space Institute, Technical University of Denmark, Elektrovej 328, DK-2800 Kgs. Lyngby, Denmark
\and
Instituto de Astrof\'isica de Canarias, E-38200 La Laguna, Tenerife, Spain
\and
Universidad de La Laguna (ULL), Departamento de Astrof\'isica, E-38206 La Laguna, Tenerife, Spain
}
   
\date{Received 8 December, 2020; accepted 25 January, 2021}

\abstract
{Using the Hamilton \'Echelle Spectrograph at Lick Observatory, we have obtained precise radial velocities (RVs) of a sample of 373 G- and K-giant stars over more than 12 years, leading to the discovery of several single and multiple planetary systems. The RVs of the long-period ($\sim 53$ years) spectroscopic binary \textepsilon\,Cyg (HIP\,102488) are found to exhibit additional regular variations with a much shorter period ($\sim 291$ days).}
{We intend to improve the orbital solution of the \textepsilon\,Cyg system and attempt to identify the cause of the nearly periodic shorter period variations, which might be due to an additional substellar companion.}
{We used precise RV measurements of the K-giant star \textepsilon\,Cyg from Lick Observatory, in combination with a large set of RVs collected more recently with the SONG telescope, as well as archival data sets. We fit Keplerian and fully dynamical $N$-body models to the RVs in order to explore the properties of a previously known spectroscopic stellar companion and to investigate whether there is an additional planetary companion in the system. To search for long-term stable regions in the parameter space around the orbit of this putative planet, we ran a stability analysis using an $N$-body code. Furthermore, we explored the possibility of co-orbital bodies to the planet with a demodulation technique. We tested the hypothesis of \textepsilon\,Cyg being a hierarchical stellar triple by using a modified version of the $N$-body code. Alternative causes for the observed RV variations, such as stellar spots and oscillations, were examined by analyzing photometric data of the system and by comparing its properties to known variable stars with long secondary periods and heartbeat stars from the literature.
}
{Our Keplerian model characterizes the orbit of the spectroscopic binary to higher precision than achieved previously, resulting in a semi-major axis of $a = \unit[15.8]{AU}$, an eccentricity of $e = 0.93$, and a minimum mass of the secondary of $m \sin i = 0.265\,M_\sun$.
Additional short-period RV variations closely resemble the signal of a Jupiter-mass planet orbiting the evolved primary component with a period of $\unit[291]{d}$, but the period and amplitude of the putative orbit change strongly over time.
Furthermore, in our stability analysis of the system, no stable orbits could be found in a large region around the best fit. Both of these findings deem a planetary cause of the RV variations unlikely.
Most of the investigated alternative scenarios also fail to explain the observed variability convincingly. Due to its very eccentric binary orbit, it seems possible, however, that \textepsilon\,Cyg could be an extreme example of a heartbeat system.
}
{}


\maketitle

\section{Introduction}

To date, more than 4000 extrasolar planets have been confirmed, and more than 800 of these were discovered with Doppler spectroscopy, also known as the radial velocity (RV) method.\footnote{\url{https://exoplanets.nasa.gov}} While most of the discovered planets orbit main-sequence (MS) stars, the number of detections around evolved stars has also risen quickly: Since the first discovery of a planet around a giant star in 2001 \citep[][]{Frink2002}, 112 exoplanets orbiting evolved intermediate-mass stars have been published.\footnote{\url{https://www.lsw.uni-heidelberg.de/users/sreffert/giantplanets/giantplanets.php}}

Giant stars allow us to extend planet surveys to higher stellar masses, while also still being sensitive in the lower mass regime: Whereas intermediate-mass MS stars show only few and rotationally broadened absorption lines, their evolved counterparts, such as K- and G-giant stars, have many sharp spectral lines and are therefore perfect targets for RV measurements. Furthermore, the discoveries of planets around these stars help to improve our understanding of the evolution of planetary systems once the stars evolve into giants \citep{Villaver2009,Reffert2015}.

Despite the overall large number of extrasolar planets discovered to date, only a comparably small fraction of $3 \sim 4$\% has been detected in stellar binaries\footnote{\url{https://www.univie.ac.at/adg/schwarz/multiple.html}}. In the case of giant stars, this number is especially small, with only five known cases: 11\,Com \citep[whose companion falls into the brown dwarf regime,][]{Liu2008}, \textgamma\,Leo \citep{Han2010}, 91\,Aqr \citep{Mitchell2013}, 8\,UMi \citep{Lee2015}, and HD\,59686 \citep{Ortiz2016}, and only one of these (HD\,59686) is a spectroscopic, that is, rather close binary. To a large extent, this small fraction of discovered planets in binary star systems can be explained by the fact that most exoplanet surveys focus on single stars, which is unfortunate since binary systems harboring planets serve as good special cases to constrain the theory of planet formation and evolution.

In this work we study the K giant star \textepsilon\,Cyg, which has been observed spectroscopically for more than 100 years and is known to undergo large RV changes, hinting at the existence of a close stellar companion to the primary component that is not directly visible. We use our own RV measurements of the star both from the Lick and Stellar Observations Network Group (SONG) telescopes to derive a precise orbit of the spectroscopic stellar companion and to investigate whether additional short-period RV variations are caused by an S-type (i.e., circumstellar) planet in the system.

In Sect.~\ref{SecStellarProperties} we present the known properties of the \textepsilon\,Cyg system and its primary component. Sect.~\ref{SecObservations} describes the RV data sets used in this analysis. Next, in Sect.~\ref{SecRVAnalysis}, we perform the RV modeling. In Sect.~\ref{SecStabilityAnalysis} we then analyze the stability of the system using an $N$-body code. In Sect.~\ref{SecAlternativeExplanations} we investigate alternative explanations for the short-period signal. Sect.~\ref{SecSummaryConclusion} summarizes and concludes our analysis.

  \section{Stellar properties}\label{SecStellarProperties}

\begin{table}
\caption{Stellar properties of \textepsilon\,Cyg\,A}
\label{StellarProp}
\centering
\begin{tabular}{l c}
\hline\hline
Parameter      &  Value \\
\hline
Apparent magnitude $m_V$ [mag]\tablefootmark{a} &  $2.48 \,\pm\, 0.01$     \\
Luminosity $L_\star$ [L$_\sun$]\tablefootmark{b} &  $57.1 \,\substack{+0.5\\-0.4}$    \\
Color index $B-V$ [mag]\tablefootmark{a}          &  $1.04 \,\pm\, 0.01$    \\
Effective temperature $T_\mathrm{eff}$ [K]\tablefootmark{b} &  $4805 \,\substack{+16\\-14}$   \\
Surface gravity $\log g$ [cm s$^{-2}$]\tablefootmark{b}     &  $2.45 \,\substack{+0.16\\-0.05}$     \\
Metallicity [Fe/H] [dex]\tablefootmark{c}         & $-0.11 \,\pm\, 0.03$     \\
Stellar mass $M_\star$ [M$_\sun$]\tablefootmark{d}         &  $1.103 \,\pm\, 0.042$    \\
Stellar radius $R_\star$ [R$_\sun$]\tablefootmark{b}       &  $10.94 \,\substack{+0.08\\-0.13}$    \\
Parallax [mas]\tablefootmark{e}                   &  $44.86 \,\pm\, 0.12$    \\
Distance [pc]\tablefootmark{e}                    &  $22.29 \,\pm\, 0.06$     \\
Age [Gyr]\tablefootmark{b}                        & $9.62 \,\pm\, 0.12$     \\
Spectral type\tablefootmark{f}                    & K0 III   \\
Frequency of max. power [\si{\micro}\si{Hz}]\tablefootmark{d}   & $32.16 \,\pm\, 0.81$ \\
\hline
\end{tabular}
\tablefoot{
\tablefoottext{a}{\cite{Oja1993}}
\tablefoottext{b}{\cite{Stock2018}}
\tablefoottext{c}{\cite{Montes2018}}
\tablefoottext{d}{Arentoft et al. (to be published)}
\tablefoottext{e}{Hipparcos, the new reduction \citep{vanLeeuwen2007}}
\tablefoottext{f}{\cite{Keenan1989}}
}
\end{table}


\textepsilon\,Cyg\,A is a bright ($m_V = \unit[2.48]{mag}$) K0 III giant star.
Its Hipparcos parallax is $\unit[44.86 \pm 0.12]{mas}$ \citep{vanLeeuwen2007}, which puts it at a distance of $\unit[22.29 \pm 0.06]{pc}$.
It has been observed spectroscopically since the beginning of the 20th century \citep[see e.g.,][]{Campbell1906,Kustner08,McMillan92}, and has long been known to host a spectroscopic binary companion, for which \citet{Gray2015} derived an orbital period around $\unit[55.1]{yrs}$ without putting an error on that value. The spectroscopic companion has never been observed visually though, and the spectra do not show any evidence of a second set of spectral lines nor any temporal changes of line asymmetry \citep{Gray1982}. 
\citet{McMillan92} also detected short-period RV variations on a time scale on the order of a few $\unit[100]{d}$ with much smaller amplitude, which they did not analyze in detail, but simply compared them qualitatively to other examples of K-giants with fast variations. Further RV measurements of the \textepsilon\,Cyg system were performed from 1999-2010
by \citet{Gray2015}, who present a spectroscopic analysis of the binary system. 
As the formal uncertainties of these measurements are too large to clearly identify the short-period signal, \citet{Gray2015} only reference the data taken by \citet{McMillan92} and propose that the variations might be caused by stellar activity, modulated by the rotation of the star.

In addition to the spectroscopic companion, \textepsilon\,Cyg\,A has an optical companion (\textepsilon\,Cyg\,B) at a separation of 71\arcsec \citep{Montes2018}, whose Gaia DR2 parallax and distance are $\unit[4.35 \pm 0.02]{mas}$ and $\unit[229.9 \pm 1.3]{pc}$ \citep{Gaia2018}, respectively, and which is therefore not gravitationally associated. A second optical companion C has a separation of 78\arcsec{} \citep{Montes2018}, and its Gaia DR2 parallax is $\unit[45.51 \pm 0.03]{mas}$, putting it at a distance of $\unit[22.97 \pm 0.01]{pc}$ and therefore very close to \textepsilon\,Cyg\,A \citep{Gaia2018}. According to \citet{Montes2018}, it is an M4V+ star, and as its proper motion also resembles that of the primary star (\textepsilon\,Cyg\,C: $\unit[+354.62 \pm 0.04]{mas/yr}$ in r.a., $\unit[+329.18 \pm 0.05]{mas/yr}$ in declination \citep{Gaia2018}; \textepsilon\,Cyg\,A: $\unit[+355.66 \pm 0.08]{mas/yr}$ in r.a., $\unit[+330.60 \pm 0.09]{mas/yr}$ in declination \citep{vanLeeuwen2007}), it can be expected to be a very wide physical companion. We do not pay any further attention to it within this work though, as its separation from the primary at the distance of the system is at least $\unit[22.29]{pc} \cdot \unit[78]{mas} = \unit[1739]{AU}$, and any influence therefore can be neglected.

We have two independent sources for the stellar properties of \textepsilon\,Cyg\,A: The first one comes from \citet{Stock2018}, who used a Bayesian interpolation scheme for evolutionary tracks. This method provides a probability for the star to either fall onto the red giant branch (RGB) or onto the horizontal branch (HB). 
It also delivers nonsymmetrical probability density functions (PDFs), which allows us to compute asymmetric $1\sigma$ confidence intervals for each of the stellar parameters. The mode values of the PDFs are used as the most probable values. According to this method, \textepsilon\,Cyg\,A is most probably an HB star ($P = 99.5\%$), with a stellar radius of $R_\star = \unit[10.94\substack{+0.08\\-0.13}]{R_\sun}$ and a mass of $M_\star = \unit[1.21\substack{+0.46 \\-0.09}]{M_\sun}$. This mass estimate is smaller than the numbers that are usually adopted in the literature, which are closer to $\unit[2]{M_\sun}$ \citep[see e.g.,][]{Gray2015}. 
It is possible that previous authors assumed \textepsilon\,Cyg\,A to be an RGB star and thus overestimated its stellar mass. This is plausible considering the (very unlikely) RGB solution by \citet{Stock2018}, which yields a mass of $M_\star = \unit[1.46\substack{+0.43 \\-0.13}]{M_\sun}$, somewhat closer to the past estimates.\footnote{The RGB solution is not published; we thank the authors for providing us with the result.}

In order to get a second independent measurement for the stellar properties, we also performed an asteroseismic campaign on \textepsilon\,Cyg\,A with SONG, and derived its mass from the frequency of maximum power $\nu_\mathrm{max}$ of the star, using the relation as in \citet{Stello2017}. We used the same approach as in \citet{Arentoft2019}, who analyzed a similar SONG-dataset for the red giant \textepsilon\,Tau, to derive $\nu_\mathrm{max}$ and its uncertainty. In short, following the methods described in \citet{Mosser2009} and \citet{Stello2017}, we applied a Gaussian fit combined with a linear (background) trend to the oscillation signal in the power spectrum to determine $\nu_\mathrm{max}$. The uncertainty was found by performing the same fit to slightly modified versions of the power spectrum, in each of which a single oscillation mode had been subtracted, and using the variations in the determined $\nu_\mathrm{max}$-values to estimate the uncertainty; see
\citet{Arentoft2019} for details. The analysis  resulted in a $\nu_\mathrm{max} = \unit[32.16 \pm 0.81]{\si{\micro}\si{Hz}}$, which gives a mass estimate of $M_\star = \unit[1.103 \pm 0.042]{M_\sun}$; the $1\sigma$-error bars of this result overlap with the ones of the HB mass estimate from \citet{Stock2018}, and we can rule out the possibility of \textepsilon\,Cyg\,A being an RGB star with high certainty. Given the robustness and high fidelity of the asteroseismic method, we adopt that mass estimate for our analysis. Table~\ref{StellarProp} lists all the stellar parameters.

\section{Observations}\label{SecObservations}

From 1999 to 2012, we monitored a sample of 373 G- and K-giant stars and measured their RV variations, using the Hamilton \'Echelle Spectrometer at the Lick observatory in California, USA. The spectrograph was fed by the 0.6\,m Coud\'e Auxiliary Telescope (CAT) and covers a wavelength range of 3755 - 9590\,\AA{} at a resolving power of $R \sim 60\,000$. High RV precision is achieved using the iodine cell method as described by \citet{Butler1996}. Our observations led to several planet detections \citep[e.g.,][]{Frink2002, Reffert2006, Trifonov2014}, among them a massive circumprimary planet in a close and eccentric binary system \citep{Ortiz2016}. The measurements also proved valuable in gaining a statistical understanding of the properties of extrasolar planets around giant stars \citep{Reffert2015}.

The K-giant star \textepsilon\,Cyg (HIP\,102488) is a member of our sample, and we collected a total of 109 RV measurements at Lick observatory. They cover a time span of more than eleven years (from June 2000 until November 2011), and their median measurement precision is $\unit[4.8]{m\, s^{-1}}$.

\begin{figure}
\centering
\includegraphics[width=\hsize]{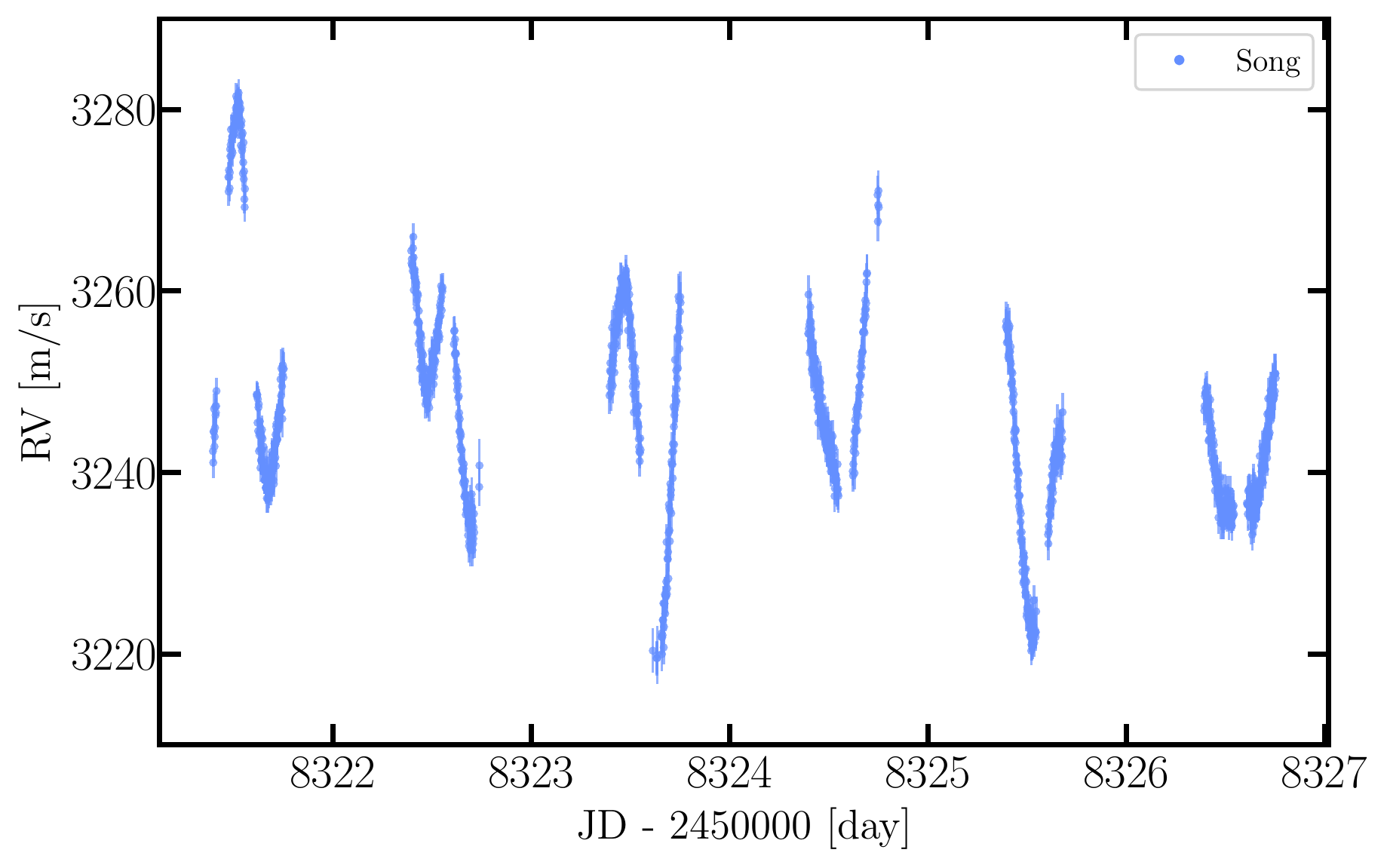}
\caption{Uncorrected SONG RV measurements from the last week of our asteroseismic campaign on \textepsilon\,Cyg. Oscillations with periods shorter than a day are clearly visible.}
\label{AsteroseismicRVs}
\end{figure}

In addition to the Lick data set, we obtained a total of $5\,272$ RV measurements of \textepsilon\,Cyg with the 1\,m robotic SONG telescope on the island of Tenerife, Spain \citep{Andersen2014,Andersen2019}. A set of $5\,063$ of these individual measurements were part of an asteroseismic campaign to better constrain the stellar properties of the K giant (see Sect.~\ref{SecStellarProperties}); these observations were therefore performed at very high cadence, with about 100 to 150 measurements per night, for a little more than a month from June 20 until July 27, 2018. From these measurements we can get an estimate of the short-term RV jitter, that is, the expected stochastic astrophysical stellar RV variation. Data from a few nights of this campaign are shown in Fig.~\ref{AsteroseismicRVs}; the stellar oscillations are clearly visible, with peak-to-peak variations of $\unit[\sim 40-50]{m\, s^{-1}}$ on timescales of several hours, in agreement with the value for $\nu_\mathrm{max}$ found above. The observed oscillations are a combination of a number of individual oscillation modes, which together give rise to a short-term variation in the time series with a standard deviation of $\unit[\sim 11]{m\, s^{-1}}$. This scatter agrees well with other stars with similar maximum-power frequencies $\nu_\mathrm{max}$ as \textepsilon\,Cyg \citep[compare][]{Yu2018}. Following \citet{Kjeldsen2008}, we isolated the p-mode signal and determined the amplitude per radial mode to be $\unit[1.45 \pm 0.06]{m\, s^{-1}}$. That value is approximately 9 times larger than the amplitude of the oscillation signal seen in the Sun, and accordingly the standard deviation and peak-to-peak variations of our asteroseismic measurements are roughly one order of magnitude larger than the solar values \citep[see][]{Kjeldsen2008, Andersen2019}.

For our orbital analysis of the system this large number of data points is unsuitable, as it puts too much weight on the SONG measurements and requires too much computational power when performing orbital fits. For this reason, we computed the median RV for each night with more than one observation. Our updated SONG data set then consists of 228 RV measurements, which fall between April 2015 and December 2018. They cover exactly the last periastron passage of the eccentric stellar companion to \textepsilon\,Cyg, which occurred around February 2017, and thus enable us to determine the binary orbit with very high precision. Just as the Lick measurements, the high-resolution spectra taken by SONG are calibrated by the means of an iodine cell \citep{Grundahl2017}, and the measurements reach a similar precision, with a median of $\unit[2.3]{m\, s^{-1}}$.

In addition to the long-period signal induced by the close stellar companion, both the Lick and SONG data sets show RV variations with periods just below $\unit[300]{d}$ (see \ref{SecGLSPeriodogram}), which could be caused by a planetary companion orbiting the primary star in an S-type configuration.
To gain a better coverage of the orbits of the binary and the possible planetary companion, we complement our own two data sets with the older measurements from \citet{McMillan92}. The McMillan data set consists of 213 individual RV measurements of \textepsilon\,Cyg, collected between May 1987 and May 1992 with an interferometer at the 0.9\,m-telescope of the Steward Observatory at Kitt Peak, USA. Calibrations were done with emission lines of a Fe-Ar hollow cathode lamp, yielding a precision of $\unit[12]{m\, s^{-1}}$ for all individual measurements.

\begin{figure*}
\centering
\includegraphics[width=\hsize]{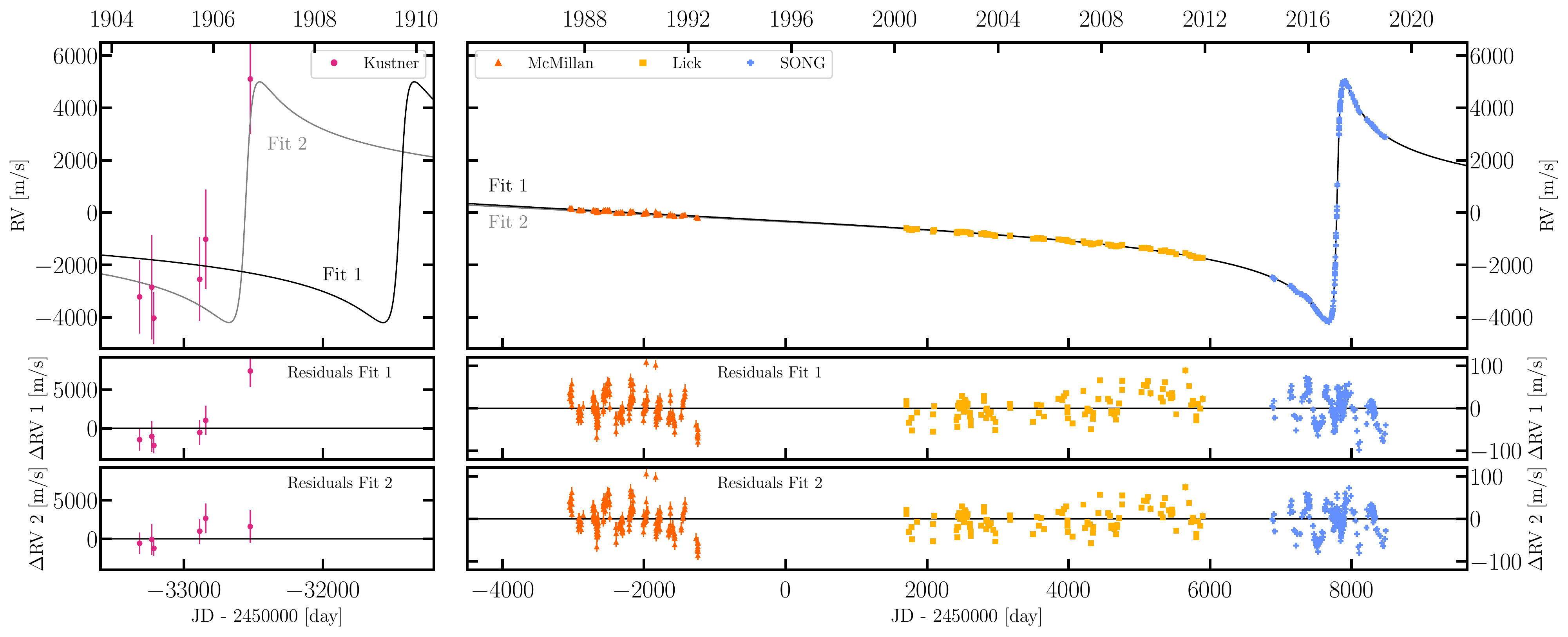}
\caption{\textit{Top plots:} RV measurements of \textepsilon\,Cyg by \citet{Kustner08} (left), and \citet{McMillan92}, Lick observatory and the SONG telescope (right), along with two orbital models of the binary companion: \textbf{Fit 1} (black line) is the best-fit solution for the three modern data sets of McMillan, Lick and SONG, with an orbital period of $\unit[19\,618.6]{d}$; \textbf{Fit 2} (gray line) was derived from fitting all four data sets and yields a period of $\unit[20\,175]{d}$. \textit{Middle plots:} Residuals of the RV measurements after removing the orbital solution of Fit 1. The last measurement of Kustner then is a clear outlier. \textit{Bottom:} Residuals of the RV measurements after removing the orbital solution of Fit 2. Now all the Kustner RVs agree with the model within their 1$\sigma$-uncertainties.}
\label{RVPlotBin}
\end{figure*}

Finally, we use six RV measurements from \citet{Kustner08} to cross-check our results for the orbital solution of the long-period stellar companion to \textepsilon\,Cyg. The observations by Kustner were carried out in Bonn, Germany, with a 30\,cm refractor and a spectrograph constructed out of three prisms, and RVs were computed by comparing the positions of the stellar absorption lines to a reference iron spectrum. The observations of \textepsilon\,Cyg cover the time between July 1904 and September 1906. Their errors are large in comparison to modern values, and vary between $1000$ and $\unit[2100]{m\, s^{-1}}$, so we do not include these measurements in our fits. The merit of this data set however lies in the fact that it seems to record the periastron passage of the stellar companion two full periods back, and we can overplot and check our results.

We decided to refrain from using the RVs collected by \citet{Gray2015} from 2001 until 2010 as they cover mostly the same time as our own Lick data and have much larger measurement errors, which is why they do not help to constrain our models. All in all, our combined data set for the fitting procedures consists of 550 RV measurements from McMillan, Lick and SONG.

\section{Analysis of the RV data}\label{SecRVAnalysis}
\subsection{Determining the orbit of the close stellar companion}\label{SecOrbitStellarCompanion}

We use a Keplerian model in combination with a Levenberg-Marquardt (LM) minimization scheme to fit the combined RV measurements of \textepsilon\,Cyg and determine orbital parameters for the spectroscopic companion, which we hereafter also refer to as the close stellar companion; the main host star is denoted as in the literature, \textepsilon\,Cyg\,A. The Keplerian fit incorporates eight free parameters: RV semi-amplitude $K$, orbital period $P$, eccentricity $e$, argument of periastron $\omega$, and mean anomaly $M_0$ of the orbit, as well as zero-point offsets for the McMillan, Lick and SONG data sets, respectively. In order to account for the RV jitter of the star, we quadratically added a fixed value of $\unit[20]{m\, s^{-1}}$ to all individual measurement errors (see Sect.~\ref{SecOrbitBothCompanions}). Error estimation was performed with a Markov-Chain Monte Carlo (MCMC) approach, using the \texttt{emcee} python package \citep{Foreman-Mackey2013}: We constructed 32 walkers (four times the number of free parameters) with initial values drawn randomly from a small Gaussian ball around the best-fit solution from above, and let the sampler run for altogether 6000 steps (where we discarded the first 1000 steps). This method produced $36\,649$ unique samples of the parameters; the $1 \sigma$ intervals of these parameter sets around the mean values of the posterior distribution serve as our uncertainties.

The LM fitting scheme delivers a best-fit model with an orbital period of $\unit[19\,619]{d}$ at $\chi^2_\mathrm{red} = 1.93$, which lies within the errors from the mean period of the MCMC posterior distribution of $\unit[19\,611.5\,\substack{+117.4\\-119.9}]{d} \approx \unit[53.7 \pm 0.3]{yrs}$ (see column ``Single-Keplerian'' of Table~\ref{TableKepPar}). Our results therefore are roughly one-and-a-half years less than the orbital period of about $\unit[55.1]{yrs}$ derived by \citet{Gray2015}, who used 53 RV measurements taken between 2001 and 2011 and complemented them with the McMillan data set as well as data by \citet{Griffin94}.

When plotting our best-fit solution with a period of $\unit[19\,619]{d}$ over the measurements taken by \citet{Kustner08}, one clearly notices that the last RV from that data set falls far away from the curve, by a value of about $\unit[7400]{m\, s^{-1}}$, which corresponds to more than three times the formal error of $\unit[2100]{m\, s^{-1}}$ (compare Fig.~\ref{RVPlotBin}). If we fit all four data sets combined, the best Keplerian model delivers an orbital period of $\unit[20\,175]{d}$ and a $\chi^2_\mathrm{red}$ of $1.95$. In this model, the former outlier now also agrees with the fit curve within its 1$\sigma$-uncertainty and falls exactly at the sharp peak of the curve; however, the fit leads to systematic (albeit small) deviations from the RVs of the three modern data sets by McMillan, Lick and SONG. As our further analysis focuses on these measurements, we decide to adopt the model with the shorter orbital period of $\unit[19\,619]{d}$ for now. We do not exclude the possibility that the result for the orbital period of the close stellar companion might still change on the level of a few percent with future observations coming in, but given our current data it seems most probable that the last Kustner measurement is actually an outlier.

Using the primary mass of $M_\star = \unit[1.103 \pm 0.042]{M_\sun}$ and solving the posterior results of the mass equation $f(m)$ numerically for fixed inclinations $i = 90^\circ$, we derive a minimum mass of $\unit[278 \pm 7]{M_{jup}} = \unit[0.265 \pm 0.007]{M_\sun}$ for the close stellar companion.
Our estimate places it therefore well above the brown dwarf regime. As the spectroscopic companion has never been imaged directly to our knowledge, and its absorption lines are not visible in the spectra, its luminosity must be much lower than that of the primary star \textepsilon\,Cyg\,A. If we additionally assume that both stars have been formed around the same time, as is expected for binaries, this opens up two possibilities for the nature of the companion: The first is a white dwarf, which would mean that it was originally the primary, more massive component in the system and therefore evolved more quickly.

The second possibility is that the spectroscopic companion to \textepsilon\,Cyg\,A is a main-sequence star with a mass lower than that of the primary, therefore evolving more slowly and being much less luminous. This places it anywhere between an early M-dwarf to an early G-type star, and constrains the possible orbital inclination to values larger than $\sim 14^\circ$, to keep its mass smaller than the mass of the primary, $\sim\unit[1.103]{M_\sun}$. Therefore the apparent brightness in the V band should be between approximately 6 and $\unit[12]{mag}$.

\citet{Griffin94} speculated about the possibility of imaging the companion directly (either at visual wavelengths or the infrared) and predicted that its angular separation on the sky could get nearly as large as 2\arcsec\ during apastron, but he had no certain solution for the orbit yet and used a semi-major axis of approximately $\unit[20]{AU}$ and eccentricity of $0.9$ for his calculations. Our orbit solution puts the apastron at $\unit[30.5]{AU}$ from the primary. At the distance of the system, $\unit[22.1]{pc}$, this would translate to a maximum projected separation of 1.38\arcsec\ in the case of an optimal position of the system on the sky. However, we were also able to derive a solid estimate of the argument of periastron of the orbit, which is at $275^\circ$ and therefore places the periastron of the orbit nearly exactly in-between the observer and the primary component, and the apastron behind the primary \textepsilon\,Cyg\,A as seen from Earth. Hence the angular separation at apastron will only become large for a low inclination of the orbit. For an inclination higher than $80^\circ$, the maximum separation of the two components would be reached around the co-vertices of the orbital ellipse (semi-minor axis: $b \approx \unit[5.8]{AU}$, projected separation: 0.26\arcsec).

\begin{figure}
\centering
\includegraphics[width=\hsize]{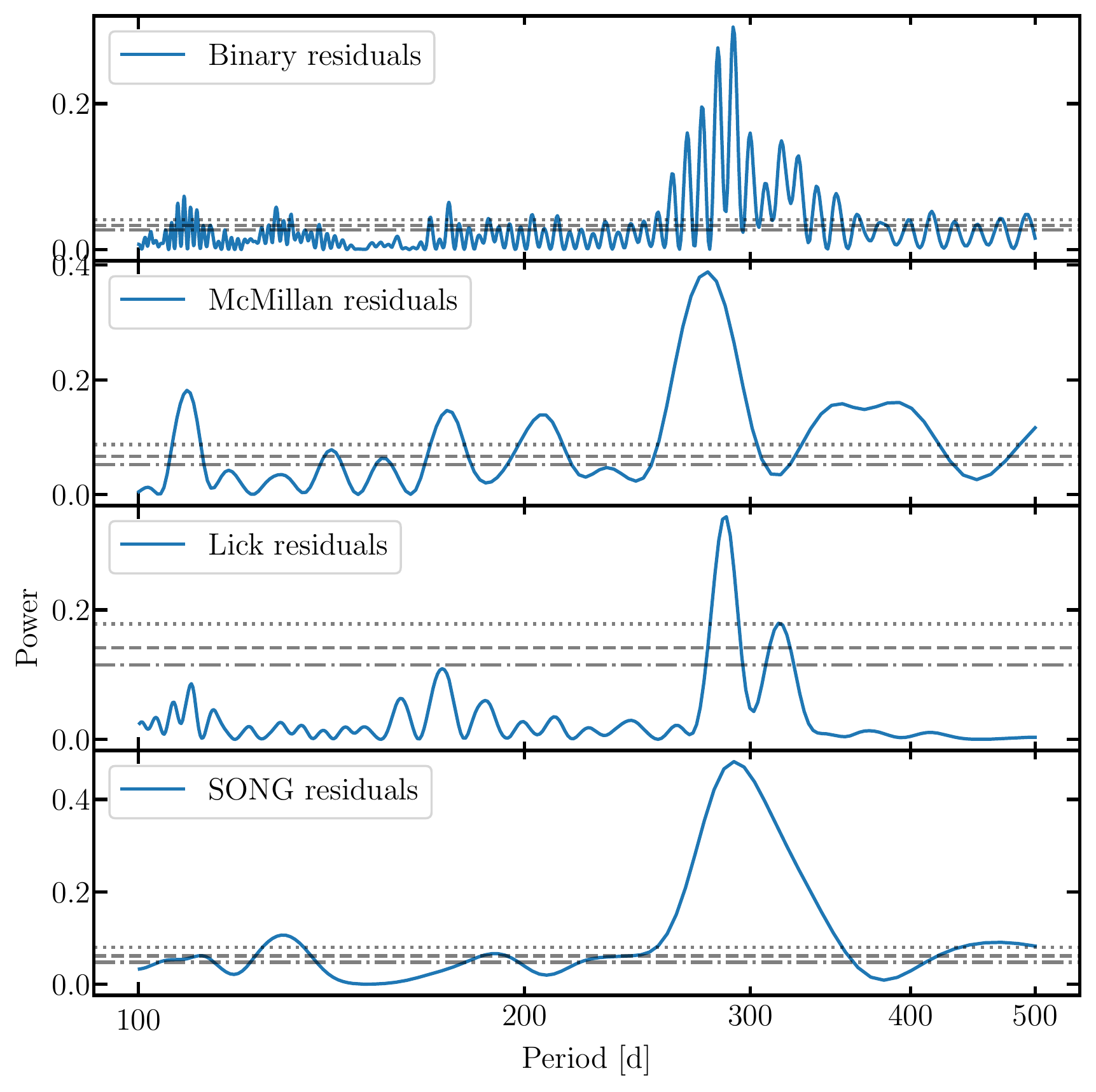}
\caption{GLS periodograms of the residuals of the RV measurements after removal of the binary signal. The periodogram of the combined data sets (top panel) shows two peaks with similar heights at $\sim\!\!\unit[282.5]{d}$ and $\sim\!\!\unit[290.8]{d}$. The gray dash-dotted, dashed and dotted lines represent false-alarm probability (FAP) levels of $5$, $1$ and $0.1\%$, respectively.}\label{GLSBinaryRes}
\end{figure}

Modern direct imaging instruments offer the capability to resolve the two stars: As an example we used the exposure time calculator of the ESO instrument \textit{SPHERE-IRDIS}\footnote{\url{http://www.eso.org/observing/etc/bin/gen/form?INS.NAME=SPHERE+INS.MODE=IRDIS}} to calculate the achievable contrast for \textepsilon\,Cyg within an exposure time of $\unit[1800]{s}$ with median weather conditions, which resulted in a magnitude difference of $\sim\!\!\unit[8.4]{mag}$ at a separation of 0.1\arcsec, and $\sim\!\!\unit[14]{mag}$ at the co-vertex separation of 0.25\arcsec. With the magnitude difference of the two stars probably not being higher than $\unit[12]{mag}$ in the visual, it might be an interesting option to try and image the spectroscopic companion directly. According to our calculations, it will reach the co-vertex around JD 2459797, that is, in August 2022, which is also the best time within the year to observe \textepsilon\,Cyg directly from Cerro Paranal Observatory.

\begin{table*}
\caption{Keplerian parameters of the \textepsilon\,Cyg binary system, from the single- and double-Keplerian models}
\label{TableKepPar}
\centering
\begin{tabular}{l | c c | c c | c c}
\hline\hline
               & \multicolumn{2}{c|}{Single-Keplerian} & \multicolumn{4}{c}{Double-Keplerian} \\
               & \multicolumn{2}{c|}{Stellar comp.} & \multicolumn{2}{c}{Stellar comp.} & \multicolumn{2}{c}{Planetary comp.} \\
Parameter      & MCMC & Best-fit & MCMC & Best-fit & MCMC & Best-fit \\
\hline
$P$ [days]     & $19611.5 \,\substack{+117.4\\-119.9}$ & $19618.6$ & $19502.9 \,\substack{+90.5\\-88.6}$ & $19575.7$ & $291.1 \,\pm\, 0.1$ & $291.2$ \\
$M_0$ [deg]\tablefootmark{a} & $160.8 \,\pm\, 1.2$ & $160.9$ & $159.7 \,\pm\, 0.9$ & $160.4$ & $138.8 \,\substack{+17.1\\-17.7}$ & $129.2$ \\
$e$            & $0.9295 \,\pm\, 0.0003$ & $0.9295$ & $0.9295 \,\pm\, 0.0002$ & $0.9297$ & $0.150 \,\substack{+0.056\\-0.058}$ & $0.173$ \\
$\omega$ [deg] & $275.30 \,\pm\, 0.06$ & $275.29$ & $275.36 \,\pm\, 0.06$ & $275.38$ & $267.04 \,\substack{+17.88\\-16.75}$ & $276.24$ \\
$K$ [$\unit{m \,s^{-1}}$]  & $4600.7 \,\pm\, 1.7$ & $4600.9$ & $4607.3 \,\pm\, 1.8$ & $4607.5$ & $29.7 \,\substack{+1.5\\-1.6}$ & $30.3$ \\
$f(m)$ [$\unit{M_{jup}}$]\tablefootmark{b} & $10.411 \,\pm\, 0.016$ & $10.407$ & $10.390 \,\pm\, 0.016$ & $10.386$ & $\left(8.009 \,\substack{+1.178\\-1.230}\right)\cdot 10^{-7}$ & $8.366 \cdot 10^{-7}$ \\
\hline
$m \sin i$ [$\unit{M_{jup}}$]\tablefootmark{c}  & $278 \,\pm\, 7$ & $278$ & $277 \,\substack{+7\\-6}$ & $277$ & $1.02 \,\pm\, 0.06$ & $1.04$ \\
$a$ [AU]       & $15.8 \,\pm\, 0.2$ & $15.8$ & $15.7 \,\pm\, 0.2$ & $15.8$ & $0.89 \,\pm\, 0.01$ & $0.89$ \\
\hline
\end{tabular}
\tablefoot{
\tablefoottext{a}{The mean anomalies are calculated at the first observational epoch in the data set of McMillan, $t_0 = 2\,446\,945.9465 \,\mathrm{JD}$.}\\
\tablefoottext{b}{The expression $f(m)$ denotes the mass function: $\frac{(m \sin i)^3}{(M_* + m)^2} = \frac{P}{2 \pi G} K^3 \sqrt{(1-e^2)^3}$.}\\
\tablefoottext{c}{Masses and errors have been derived by solving the mass function numerically for fixed inclinations $i$ of $90^\circ$, taking the uncertainty of the primary mass $M_*$ into account.}
}
\end{table*}


\subsection{GLS periodogram of the RV residuals}\label{SecGLSPeriodogram}

We aim to find relevant short-period variations in the RV measurements of \textepsilon\,Cyg by calculating generalized Lomb-Scargle (GLS) periodograms as described in \citet{Zechmeister2009}. As the RV data are largely dominated by the signal of the stellar companion with a semi-amplitude of $K \approx \unit[4.6]{km\,s^{-1}}$, we first subtracted the best-fit orbital solution of the binary from the measurements and then calculated the periodograms of the residuals. Figure~\ref{GLSBinaryRes} shows the significance of variations with periods between $100$ and $\unit[500]{d}$, for the three individual data sets by McMillan, Lick, and SONG (lower three panels), as well as for the combined measurements (uppermost panel). False-alarm probabilities (FAPs) of $5$, $1$, and $0.1\%$ are depicted by the gray dash-dotted, dashed, and dotted lines, respectively.

All three data sets show a highly significant peak at periods slightly shorter than $\unit[300]{d}$ that greatly exceeds the FAP of $\unit[0.1]{\%}$, but the exact positions of those peaks vary: In the McMillan data the highest peak lies at $\unit[277.9 \pm 2.8]{d}$, the Lick data set shows its strongest signal at $\unit[287.3 \pm 1.7]{d}$, and the SONG data set at $\unit[291.2 \pm 3.0]{d}$. The uncertainty estimation used by \citet{Zechmeister2009} is based on the curvature of the peaks in the periodograms \citep[as described in][]{Ivezic2014}, and in simulations, using fixed periods with varying white noise amplitudes and the sampling of the original data sets, we found a good agreement with these errors.

The period of the most dominant signal changes by more than $\unit[13]{d}$ over the course of 30 years from the first until the last measurement. All the peak periods lie at least 1$\sigma$ away from each other; in the case of the McMillan and Lick periods the separation is even larger than 3$\sigma$. The change of the period over time therefore seems to be quite significant, and we examine that behavior more closely in Sect.~\ref{SecTemporalEvolution}.

The GLS periodogram of the combined data sets shows a forest of peaks around periods of $\unit[300]{d}$, with the strongest one at $\unit[291]{d}$. The side peaks can at least partly be attributed to the sampling of the data: Between the individual data sets there are gaps of several years without any data, which can produce such patterns \citep[see e.g.,][]{VanderPlas2018}.

\subsection{Fitting a double-Keplerian model to the data}\label{SecOrbitBothCompanions}

We first investigate the hypothesis that this short-period signal is caused by a possible additional companion of planetary nature in the system orbiting the main component. Therefore we use a double-Keplerian model to constrain the orbits of the known stellar component and the possible second companion simultaneously. The fit then incorporates 13 free parameters: RV semi-amplitude $K$, orbital period $P$, eccentricity $e$, argument of periastron $\omega$, and mean anomaly $M_0$ for each of the companions, as well as zero-point offsets for the McMillan, Lick and SONG data sets, respectively. Again we estimated the errors through the MCMC method, similar to the single Keplerian model, but now with 52 walkers. The $1 \sigma$-intervals of the $62\,535$ samples around the mean values of the posterior distribution serve as our uncertainties.

The best-fit results from the model and the mean orbital parameters from the MCMC posterior distribution are shown in Table~\ref{TableKepPar} (columns ``Double-Keplerian''), and they agree within the errors. The parameters of the stellar companion only change slightly as compared to the single-Keplerian model. The best-fit orbital period of the putative planet is $\unit[291.2]{d}$, and it has a low eccentricity of $\sim 0.173$. Using the primary mass of $M_\star = \unit[1.103]{M_\sun}$ puts the minimum mass of the planet at about $\unit[1]{M_{jup}}$, and its semi-major axis at $\unit[0.89]{AU}$.

\begin{figure}
\centering
\includegraphics[width=\hsize]{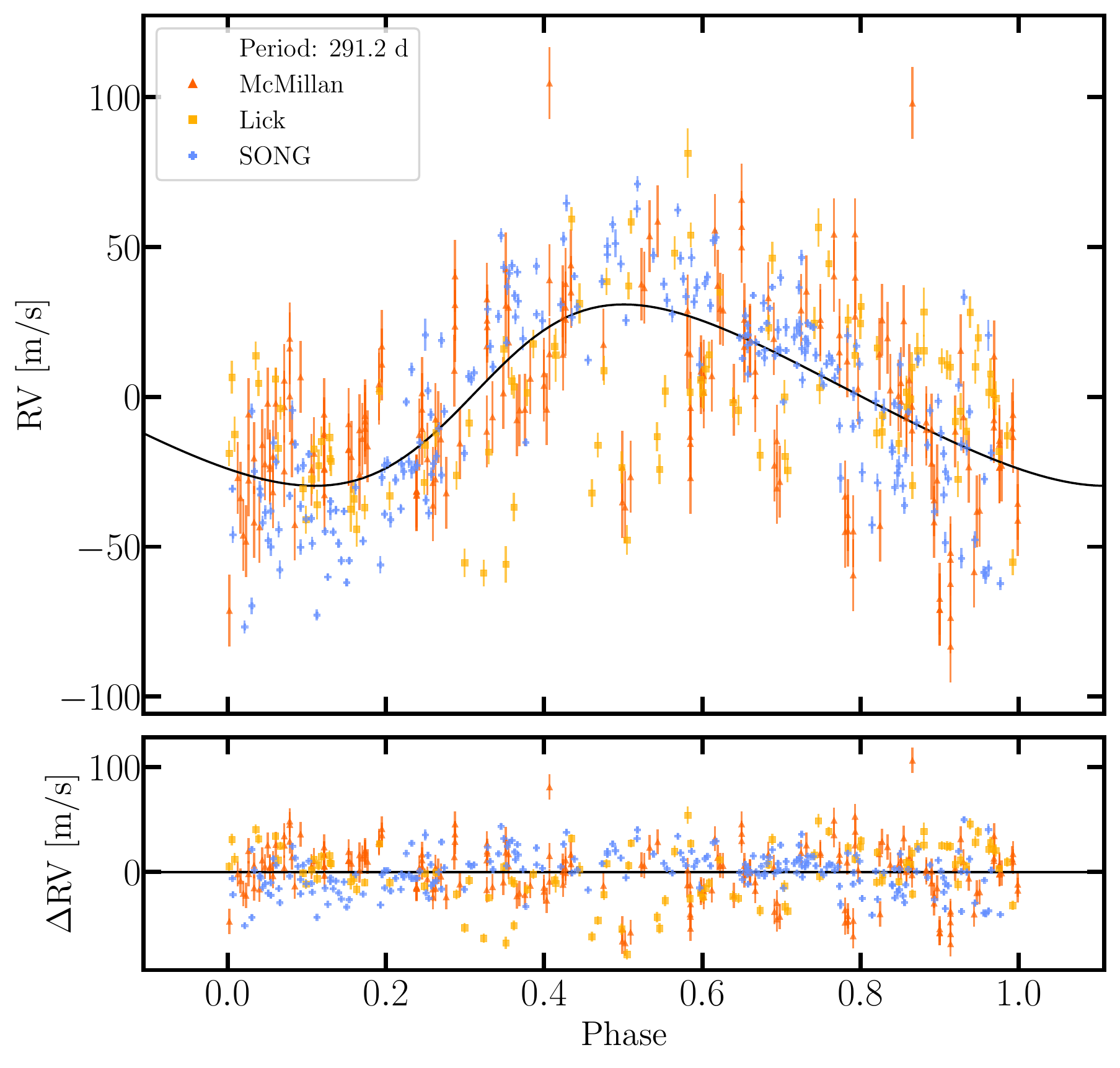}
\caption{\textit{Top panel}: RVs of the three modern data sets McMillan, Lick and SONG, phase-folded by the best-fit period of the putative planet $P = \unit[291.2]{d}$, are shown in symbols along with the RV curve of the best fit as a black line. \textit{Bottom panel}: Residuals of the individual data points.}\label{RVphasedplanet}
\end{figure}

Without taking any jitter into account, the $\chi^2_\mathrm{red}$ of the best fit is $37.49$. When we fit our Lick data set alone, letting all parameters vary, a jitter value of $\unit[17.4]{m\, s^{-1}}$ brings the $\chi^2_\mathrm{red}$ of that fit down to unity. This value is larger than the jitter estimate from the asteroseismic measurements (see Sect.~\ref{SecObservations}), which is no surprise as the Lick data set covers a much longer baseline. Stellar jitter stems from a number of stellar phenomena, and oscillations are just one of those -- others are granulation, stellar rotation, and magnetic cycles \citep{Dumusque2016, Dumusque2017}. These jitter sources act on considerably different time scales (few hours to several years), which might well explain the discrepancy between the jitter estimates for the whole Lick data set (covering ~11 years) and the asteroseismic data ($\sim$1 month). For consistency, similarly as for the Lick data we also fitted only the SONG measurements, but left out the asteroseismic data, which resulted in a jitter estimate of $\unit[15.0]{m\, s^{-1}}$ to achieve $\chi^2_\mathrm{red} = 1$. This value falls in-between the jitter estimates from the longer-baseline Lick data and the shorter-baseline asteroseismic measurements.

We can also compare these results to the color-dependent distribution of jitter estimates found for the stars in our own K-giant sample \citep[see][]{Frink2001,Trifonov2014}: For the color of \textepsilon\,Cyg, $(B-V) = \unit[(1.04 \pm 0.01)]{mag}$, the measured values scatter from about 10 to $\unit[30]{m \,s^{-1}}$, so all of the jitter results from above lie within the distribution. In the end we decided to use a fixed jitter value of $\unit[20]{m \,s^{-1}}$ for all our Keplerian and dynamical fits in this analysis. This corresponds roughly to the mean of the above-mentioned jitter distribution for stars of similar color $(B-V)$, and allows to account for additional jitter sources that only come into play over the 32-years long baseline of the three datasets McMillan, Lick and SONG combined. The double-Keplerian fit to all these measurements then delivers a $\chi^2_\mathrm{red}$ of $1.21$, which is a clear improvement to the single-Keplerian model with $\chi^2_\mathrm{red} = 1.93$ from Sect.~\ref{SecOrbitStellarCompanion}. The combined rms of all data sets is $\unit[23.85]{m\, s^{-1}}$, which corresponds roughly to our adopted jitter value.

Figure~\ref{RVphasedplanet} shows the RVs of the three data sets along with the best Keplerian fit to the data phased to the best-fit period $P = \unit[291.2]{d}$ (top) and the residuals of the RVs from the fit (bottom). While many of the data points lie close to the best-fit solution, there are a number of McMillan and Lick measurements that fall far away from the curve. They form an additional ``branch''; they are systematically shifted in phase and offset in vertical direction with respect to the fit. The amplitude of the best-fit curve is a little too small to perfectly match the distribution of the SONG RVs, but overall the SONG data set is much better represented by the model than the other two. The fact that the SONG data dominate the fit is not surprising as this is the largest data set with the smallest measurement uncertainties. However, the outliers especially in the McMillan and Lick data sets show that the double-Keplerian model does not provide a satisfactory fit to the combined data. A possible explanation for that would be a temporally changing RV signal; as the Keplerian model assumes undisturbed orbits, changes of the Keplerian elements are not incorporated.

\subsection{Investigating the temporal evolution of the Keplerian signal}\label{SecTemporalEvolution}

\begin{figure}
\centering
\includegraphics[width=\hsize]{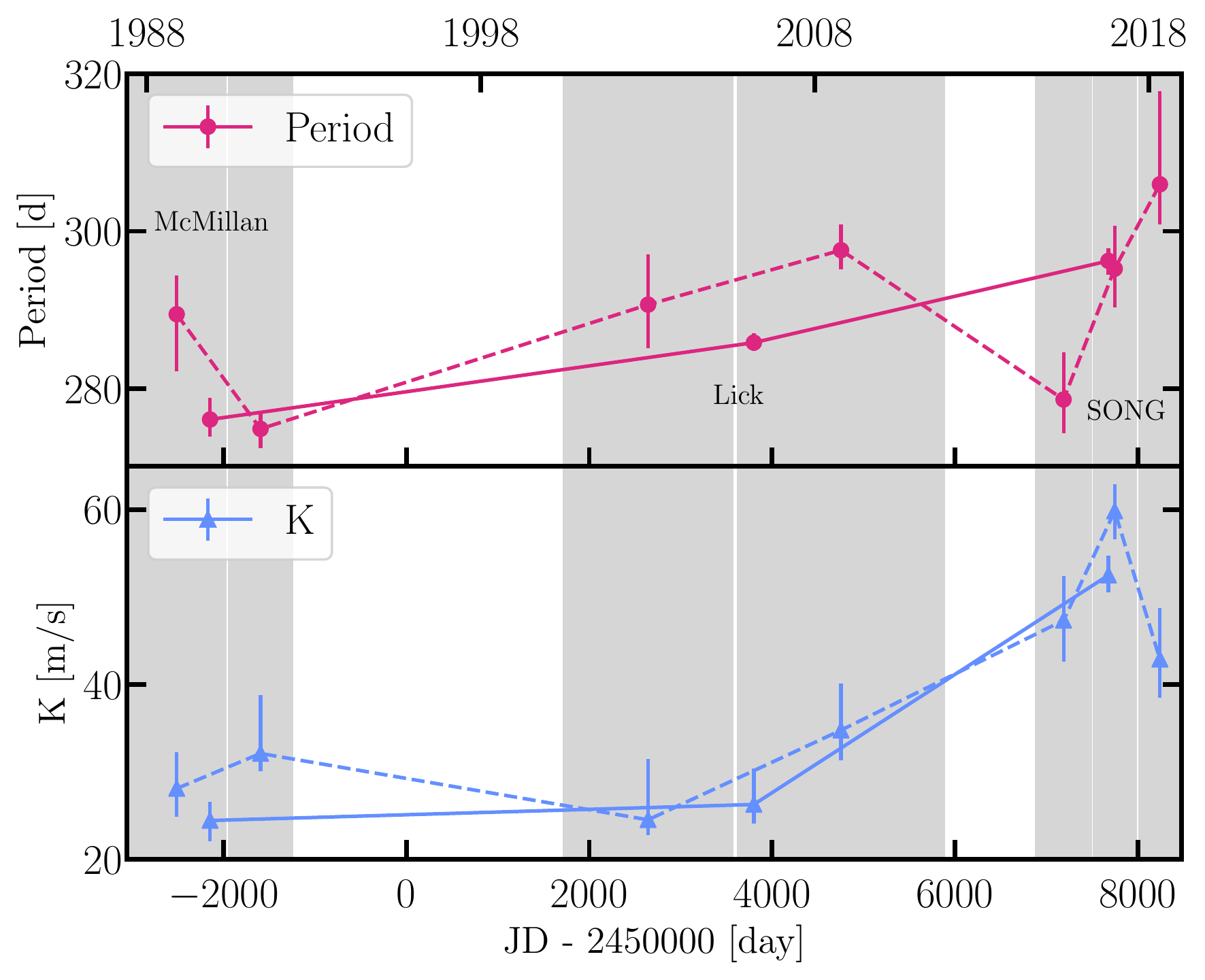}
\caption{Orbital period $P$ (top) and semi-amplitude $K$ (bottom) of the presumed planet, derived from double-Keplerian fits to the individual data sets (solid lines) and to shorter sections of the data (dashed lines). The gray-shaded areas denote the time spans from the first until the last measurement within each section.}\label{ParameterVariations}
\end{figure}

\begin{figure*}
\centering
\includegraphics[width=\hsize]{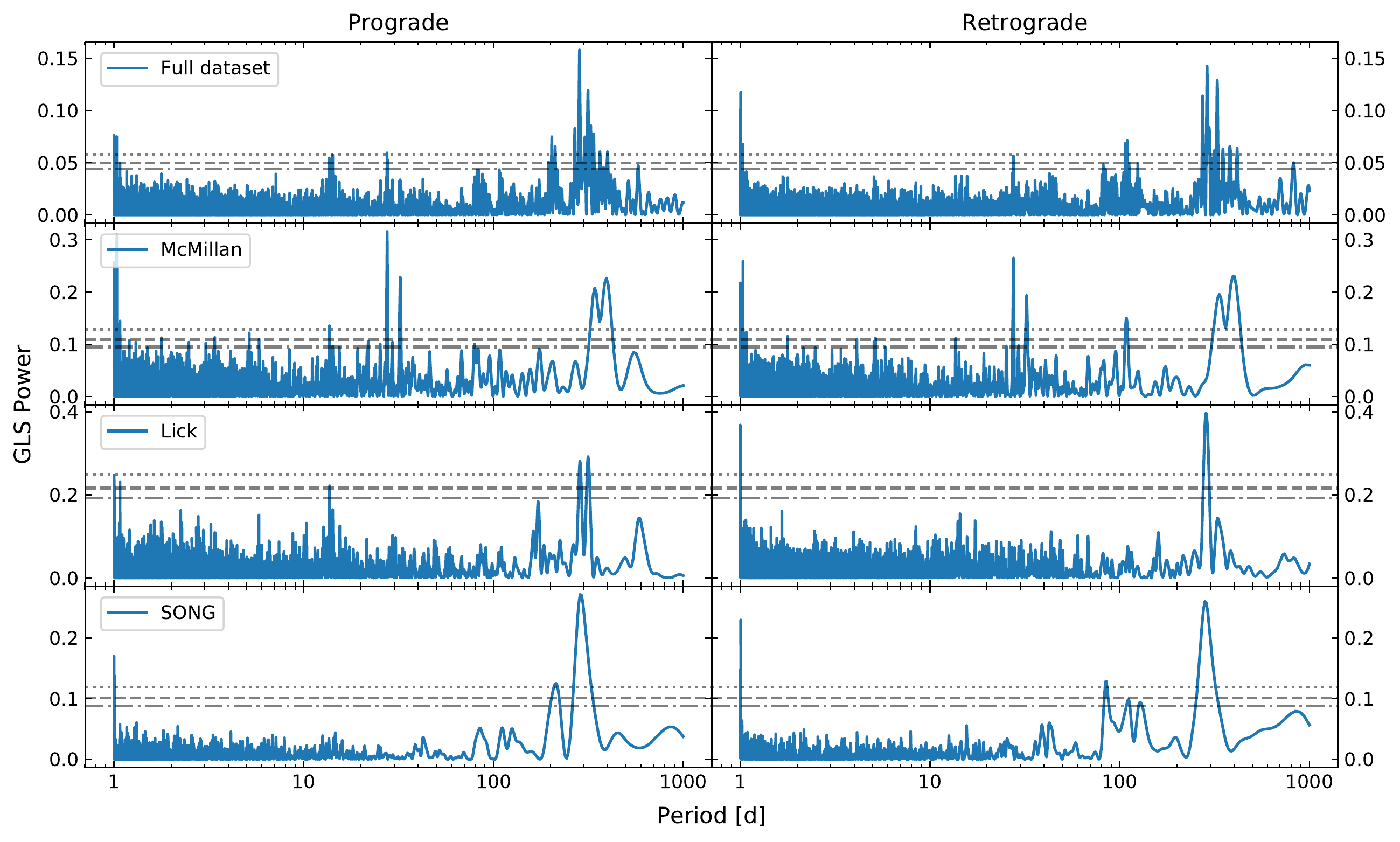}
\caption{GLS periodograms of the residuals of the best dynamical models for the prograde (left) and retrograde configuration (right). The top-most plot shows the periodograms of the residuals of the combined data, the bottom three of the individual data sets. FAPs of $5$, $1$ and $0.1\%$ are indicated by the gray dash-dotted, dashed and dotted lines, respectively.}\label{DynamicModelGLS}
\end{figure*}

To further investigate the problems of the Keplerian model, we examine the three data sets individually by fitting a double-Keplerian model to each one of them separately and comparing the resulting parameters. By this we are hoping to find out whether a changing RV signal is the cause for the bad fit of the combined model. In a second step, we split each data set into shorter sections and fit each of these separately, in order to follow changes of the Keplerian elements over even shorter times. It is important to assure a large number of data points that is more or less evenly sampled as well as a time coverage that is longer than the period of the signal for each of the sections. The McMillan data set covers 5 years with 213 measurements, but the measurement uncertainties are comparatively large ($\unit[12]{m\, s^{-1}}$), and the observations were mostly taken in chunks within a few days, leaving gaps of several days or weeks in between. This leads to a potentially poor phase coverage of the period, which is why we split the McMillan data set into only two sections of 106 and 107 data points, respectively. The Lick observations cover the longest time span of the three data sets used here, but also contain the smallest number of measurements. Therefore we also split them into two sections of 55 and 54 data points, respectively. The SONG data set in contrast consists of a large number of measurements with small uncertainties, which allowed us to split it into three sections: The first consists of 46 data points and covers the time right before the very quick RV change during periastron passage of the close stellar companion occurs; the second is made up of the 128 data points taken during the close approach of the two stars; and the third consists of 54 data points and focuses on the time after periastron passage.

Each single fit to a complete data set incorporates 11 parameters, two times five Keplerian elements for the stellar and the planetary companion, and one RV zero point. When fitting the shorter sections of the data sets however, the long-period orbit of the spectroscopic companion is not constrained sufficiently, due to the smaller number of data points and shorter time spans; therefore, in the models for the individual sections, we left the Keplerian parameters of the stellar companion fixed at the best-fit results from the whole respective data set.
This leaves 6 free parameters for each fit to a section: 5 Keplerian elements for the planetary companion and one RV zero point. Uncertainties on the free parameters were again computed from an MCMC analysis for each data set and each section, in analogy to the single and double-Keplerian models to the combined data set.

Each model to a subset of the full data, either individual data set or even shorter section, results in a much better fit than the complete double-Keplerian model, with only very few random outliers. The $\chi^2_\mathrm{red}$ for the McMillan, Lick and SONG data sets are 0.96, 0.77, and 0.47, respectively. Figure~\ref{ParameterVariations} shows the evolution of the orbital period $P$ and the semi-amplitude $K$ of the short-period signal over time: The solid lines depict best-fit results of the individual data sets, while the dashed lines denote the results from fitting the sections. In both cases distinct changes of the parameters over time become apparent: The period clearly follows the pattern already observed in the GLS periodograms of the three data sets, changing from $\unit[278]{d}$ in McMillan over $\unit[286]{d}$ in Lick to $\unit[296]{d}$ in the SONG data. The fits to the shorter sections follow a similar trend, but show more scatter. Some of that variance might be explained by the fewer measurements within each section and the shorter time spans covered by the data. Especially in the case of the SONG data set, the first and the last section only contain around 50 measurements, and none of the three sections covers two full periods of the short-period signal.

The semi-amplitude $K$ of the signal in contrast stays mostly constant for McMillan and Lick around a value of $\unit[26 \sim 27]{m\, s^{-1}}$, before it drastically increases to $\unit[53]{m\, s^{-1}}$ in the SONG data, effectively doubling. Again, the fits to the sections show some more scatter but generally follow the same trend. Interestingly, of the three SONG sections the second one, which covers the time around periastron, results in a much larger $K$ than the other two. This might hint at an actual physical process that increases the semi-amplitude exactly during periastron passage, but could also be simply explained by the fitting problems described above. According to our calculations, for the edge-on configuration the gravitational redshift $z_\mathrm{grav}$ only changes by $0.6$ to $\unit[0.7]{m\, s^{-1}}$ over the eccentric orbit of the stellar companion, so it fails as an explanation for the varying short-period semi-amplitude.

\subsection{Fitting a dynamical model to the RV data}\label{SecDynamicalModel}

One possibility to explain the changes of the Keplerian elements over time would be the gravitational interaction of the planet in question with the close stellar companion: The double-Keplerian model assumes undisturbed orbits around the center of mass and neglects any other forces that might be present. However, for two bodies in such close proximity as the planetary and stellar companion in our model, and with such high masses, this assumption certainly does not hold true anymore, and the orbits of the two bodies would be subject to considerable changes. This is even more true because our total observational time span covers more than 30 years by now, over which any evolution of the orbits can be traced. Furthermore, the SONG measurements were taken around the periastron passage of the close stellar companion, where all three bodies come closest to each other and the biggest changes can be expected.

To take this effect into account, we attempt to better fit the data using a fully dynamical model, as done before for HD\,59686 in \citet{Trifonov2018}. It consists of a modified version of the Bulirsch-Stoer $N$-body integrator in the \textit{SWIFT} package, that outputs the RV data of the primary component in the system, and a Levenberg-Marquardt minimization algorithm. As for the double-Keplerian model, the free parameters comprise the RV semi-amplitude $K$, orbital period $P$, eccentricity $e$, argument of periastron $\omega$, and mean anomaly $M_0$ for each of the companions, as well as zero-point offsets for the McMillan, Lick and SONG data sets, respectively.

Additionally, in a dynamical model the data can potentially constrain the inclinations of the orbits relative to the sky plane $i_\mathrm{1,2}$ and the longitudes of the ascending nodes $\Omega_\mathrm{1,2}$ of the close stellar companion (subscript 1) and the putative planet (subscript 2). However, it is not feasible to keep both parameters free in our case as there are many local minima in the log(likelihood)  plane, and we do not have any independent knowledge of the true orientation of the orbits that we could use to find suitable starting parameters. Therefore, we applied our model only to strictly coplanar prograde orbits, where $i_1 = i_2 = 90^\circ$ and $\Omega_1 = \Omega_2 = 0^\circ$, and strictly coplanar retrograde orbits, that is, $i_1 = i_2 = 90^\circ$ and $\Omega_1 = 0^\circ$, $\Omega_2 = 180^\circ$.

Our best dynamical models however show no improvements to the double-Keplerian fits, with a $\chi_\mathrm{red}^2$ of 1.42 and 1.23 for the prograde and retrograde configuration, respectively. Both models are still left with many systematic outliers from the best-fit curve, so it seems as if the changes of the Keplerian parameters over time cannot be fully accounted for by the expansion to a dynamical approach. When integrating the two models further over time, with a modified version of the Wisdom-Holman $N$-body integrator in the \textit{SWIFT} package, we find stability time of only 85 and $\unit[454]{yr}$ for the prograde and retrograde configuration, respectively, after which the planetary eccentricity becomes larger than 1 and the planet is kicked out of the system. The long-term survival of the putative planet therefore seems highly questionable, and we present a more in-depth analysis of the dynamical stability in Sect.~\ref{SecStabilityAnalysis}.

Figure~\ref{DynamicModelGLS} shows the GLS periodograms of the residuals of the dynamical models and the RVs, for both the prograde and retrograde configuration and for all individual data sets as well as the combined data. In all periodograms there are still peaks left that exceed the FAP level of $0.1 \%$ and mostly lie at periods close to the fitted orbital periods just below $\unit[300]{d}$, which illustrate the above-mentioned systematics in the residuals. It is evident that the dynamical models do not fully describe the signals present in the RV data.

\subsection{The co-orbital scenario}\label{SecCoorbital}

\begin{figure}
\centering
\includegraphics[width=\hsize]{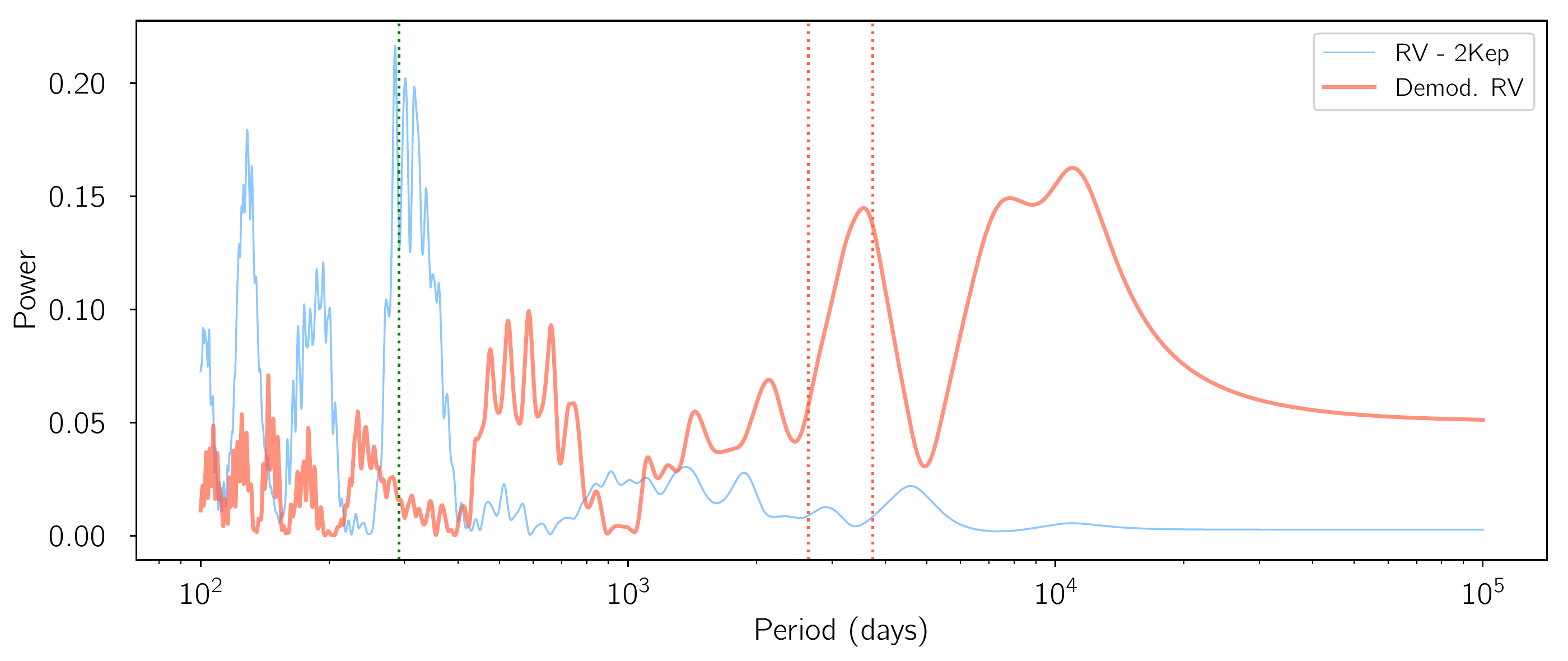}
\caption{Lomb-Scargle periodogram of the radial velocity residuals (blue) and the corresponding demodulated radial velocities (red). The red vertical dotted lines correspond to the range of allowed libration periods for this star-planet system based on a minimum ($m_t=0$) and a maximum ($m_t=m_p$) mass for the co-orbital body.}\label{fig:coorbital}
\end{figure}

In this section we investigate the source of the significant peak of the RV residuals after subtracting the possible planet signal in the context of the potential presence of a co-orbital body. Co-orbital configurations have not yet been detected outside of the Solar System although several candidates have recently been published (of particular interest is the case of TOI-178, \citealt{leleu19}; but see also \citealt{hippke15}, \citealt{janson13} or the TROY project\footnote{\url{www.troy-project.com}} by \citealt{lillo-box18a,lillo-box18b}). Interestingly, co-orbital planet pairs are stable under a very relaxed condition developed in \citet{laughlin02}, stating that such configurations would remain long-term stable as long as the total mass of the planet and its co-orbital companion is smaller than 3.8\% of the mass of the star.

The motivation of this analysis for our system comes from the theoretical study of \citet{leleu15}, who derived the implications on the RV periodogram of a co-orbital planet. In such case, the long-term libration of the co-orbital motion would introduce two signals in the periodogram at $n\pm\nu$, where $n$ is the frequency of the orbital period and $\nu$ corresponds to the libration frequency. However, the sampling and precision of the data can hide these two signals in the periodogram. \citet{leleu15} propose a technique to enhance their detection through the so-called demodulation technique:
The RV residuals are convolved with a sinusoidal function of same frequency as the carrier (the Keplerian frequency of the planet), reproducing the peak at the libration period. The libration period depends on the Trojan, planet and star masses so a minimum ($m_t=0$) and maximum ($m_t=m_p$) libration period can be estimated. We applied this technique and obtained the periodogram of the demodulated RVs, which is shown in Fig.~\ref{fig:coorbital}. The result displays a peak in the expected period range of possible libration frequencies, which opens up the possibility of a librating co-orbital body to be the cause of these RV residual peaks. A more in-depth analysis is needed to properly investigate this possibility, which is beyond the scope of this paper. It is also necessary to point out that the issue of instability of a single circumprimary planet, which we examine in depth in Sect.~\ref{SecStabilityAnalysis}, potentially also poses a problem for the co-orbital scenario; further investigations into this direction therefore require a thorough dynamical study.

\section{Dynamical stability analysis}\label{SecStabilityAnalysis}

\begin{figure*}
\centering
\includegraphics[width=\hsize]{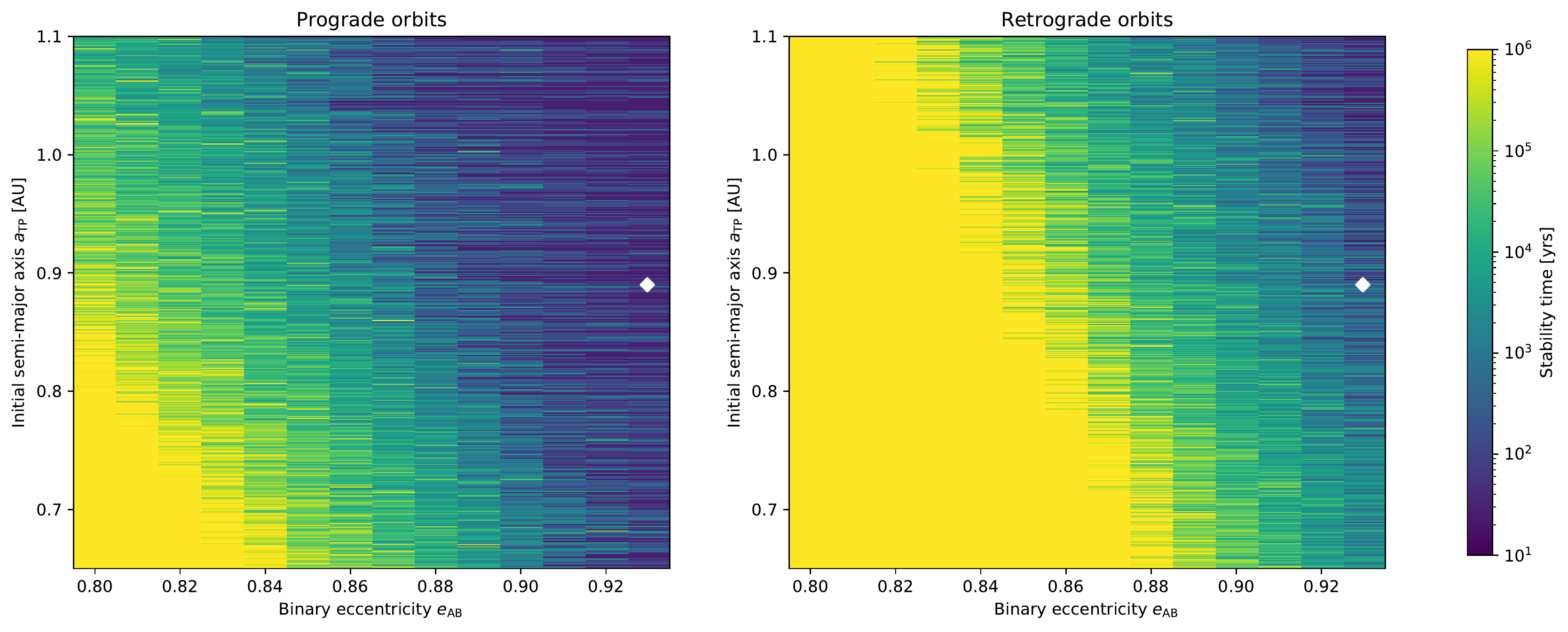}
\caption{Survival times for test particles placed into the \textepsilon\,Cyg stellar binary, in prograde (left plot) and retrograde coplanar configuration (right). The eccentricity of the orbit of the stellar companion and the initial semi-major axes of the test particles were varied in order to find the stability boundary. The position of the best-fit solution from Sect.~\ref{SecOrbitBothCompanions} is plotted by the white diamond; its size is at least by a factor of 10 larger than the uncertainties on the Keplerian parameters. }\label{StabilityTimes}
\end{figure*}

\subsection{Theoretical considerations about the orbital configuration}\label{SecTheoreticalDynamicalConsiderations}

An S-type planet around \textepsilon\,Cyg\,A would be subjected to substantial gravitational forces from the close stellar companion, which could potentially alter its orbit considerably over time, as is explained above in Sect.~\ref{SecDynamicalModel}. Even though the semi-major axis of the orbit of the stellar companion is $a_1 = \unit[15.8]{AU}$, due to its high eccentricity the periastron distance is only $q_1 = \unit[1.11]{AU}$, which is just $\unit[0.22]{AU}$ larger than the proposed semi-major axis of the planet, $a_2 = \unit[0.89]{AU}$. According to \citet{Hamilton1992}, the Hill radius of the stellar companion can be approximated as

\begin{equation}\label{Hillradius}
r_{\mathrm{H}} \approx a_1 (1 - e_1) \sqrt[3]{\frac{m_1}{3 m_0}} \approx \unit[0.267]{AU} \, ,
\end{equation}

with $m_1 = \unit[0.265]{M_\sun}$ being the minimum mass of the stellar companion, $m_0 = \unit[1.103]{M_\sun}$ the mass of the primary, and $a_1$ and $e_1$ the semi-major axis and the eccentricity of the orbit of the stellar companion. For the best coplanar double-Keplerian model the orbital path of the putative planet thus would pass through the Hill sphere of the close stellar companion during its periastron passage, making its long-term survival very questionable. For low inclinations and thus higher masses of the stellar companion, the problem becomes even worse as the Hill sphere becomes larger. However, despite the small distance of the orbits during periastron, the two companions could in principle always be much farther apart than the Hill radius of the stellar companion, as they do not necessarily have to have the same true anomalies at the time of closest approach.

We know of a number of extreme multi-companion systems today that are dynamically stable because their orbits are locked in very specific configurations, for instance, secular alignments of their periastra and/or mean-motion resonances (MMR), such as HD\,59686 \citep{Trifonov2018} or HD\,82943 \citep{Tan2013}. In order to gain a better understanding of the possible orbital configurations of the \textepsilon\,Cyg system, we ran a comprehensive dynamical analysis: Using a modified version of the Wisdom-Holman $N$-body integrator in the \textit{SWIFT} package, we tested the temporal evolution of many different orbital configurations by varying the starting parameters in a broad range around our best-fit results. First we tested both prograde and retrograde coplanar orbits, with a special focus on configurations that are locked in secular apsidal alignment or mean-motion resonance. In a second step we extended our analysis to mutually inclined orbits, for which we concentrated on regions of the parameter space near the fixed point of the Kozai-Lidov mechanism as these would offer the greatest chances for long-term stability \citep[for more details see e.g.,][]{Kozai1962,Lidov1962,Lithwick2011}.

\subsection{Stability analysis of coplanar orbits}\label{SecCoplanarOrbitsDynamics}

We set up a system of two bodies, using the derived masses of the primary and secondary stellar components as well as the semi-major axis of the secondary from Sect.~\ref{SecStellarProperties} and \ref{SecOrbitBothCompanions}, and inserted a variety of massless test particles around the best-fit orbital solution of the planet.
As starting parameters, we chose the periastron of both the test particles and the stellar companion to be at $0^\circ$ (so $\Delta\omega = 0^\circ$), and the mean anomalies of the test particles to be $0^\circ$, while that of the stellar companion was set to $180^\circ$. The semi-major axes of the test particles were varied within a range of $\unit[0.06]{AU}$ around the best-fit solution of the planet, and their eccentricities varied between 0 and 0.8. In order to understand the influence of the high binary eccentricity on the inner orbits, these simulations were repeated for eccentricities of the stellar companion ranging between 0.8 and the best-fit value of 0.93. The whole analysis was done both for prograde and retrograde orbital configurations, and a system was considered stable if the test particle stayed within a critical distance interval \{$a_\mathrm{c,1}$,\,$a_\mathrm{c,2}$\} from the primary component for the whole integration time of $10^4$ orbits of the stellar companion ($\sim \unit[5.4 \cdot 10^5]{yrs}$). The inner boundary $a_\mathrm{c,1} = \unit[0.22]{AU}$ was chosen to be the distance at which one orbit of a test particle is only resolved by 10 integration time steps (the time step of the simulation being $\unit[3.4]{d}$), the outer boundary $a_\mathrm{c,2}$ corresponds to the semi-major axis of the stellar companion.

As was expected from our considerations in Sect.~\ref{SecTheoreticalDynamicalConsiderations}, the system generally shows a very high degree of chaotic behavior. The most stable coplanar configurations were achieved for retrograde MMR or secularly aligned orbits with test particle eccentricities around 0.4. A 1:$n$ MMR is established for systems where at least one of the MMR angles
\begin{equation}
\lambda_2 - n \lambda_1 + (m-1) \omega_2 - (m-n) \omega_1 ,
\end{equation}
librates around a constant, with $\lambda_{1,2}$ being the mean longitudes of the stellar companion and the test particle, respectively. For prograde orbit, $n$ is positive and $m = 1, \ldots, n$. For retrograde orbit, $n$ is negative and $m = 1, \ldots, |n| + 2$. For secular alignment the angle between the pericenters of the two orbits,
\begin{equation}\label{SecAlignmentequation}
\Delta \omega = \omega_1 - \omega_2  \, ,
\end{equation}
librates around zero. For all our simulated systems we checked whether these requirements were fulfilled; we were thus able to identify the MMR and secularly aligned configurations. Nevertheless, even for those configurations general long-term stability only occurred for eccentricities of the stellar companion being smaller than 0.84 (for prograde orbits) or smaller than 0.9 (for retrograde ones). At the best-fit binary eccentricity of $0.93$ the instability timescale is on the order of only 10 binary periods ($\cong \unit[500]{yrs}$) in the retrograde case and on the order of just 1 binary period in the prograde case. Altering the semi-major axes of the test particles within the range defined above has barely any effect on the stability.

We redid the simulations for 1000 test particles with initial semi-major axes evenly distributed between 0.65 and $\unit[1.1]{AU}$ and fixed eccentricities at 0.1, while varying the binary eccentricity between 0.8 and 0.93 in steps of 0.01, to illustrate our general findings for the pro- and retrograde case in Fig.~\ref{StabilityTimes}. Here it becomes clear that orbiting closer-in to the primary component improves the stability of the test particles, but the position of the best fit (indicated by the white diamond) is far away from the stable region even in the retrograde configuration. With the eccentricity of the orbit of the stellar companion being our best-constrained parameter ($e_\mathrm{AB} = 0.9295 \,\pm\, 0.0003$ in the single-Keplerian model), there is no doubt that the putative planet falls into a highly unstable region of the parameter space.

\subsection{Stability analysis of mutually inclined orbits}\label{SecMutuallyInclinedDynamics}

From our considerations in Sect.~\ref{SecOrbitStellarCompanion} we know that the inclination of the binary orbit must be larger than $14^\circ$, but apart from that we do not have any constraints on the orbital inclinations. Therefore, in order to test the stability of mutually inclined orbits, we decided to focus on the region of possible Kozai librations within the parameter space, which we deem most promising to guarantee long-term stability. We ran simulations for retrograde orbits only (these being more stable than prograde orbits), varying the mutual inclination $\Delta i$ between $90^\circ$ and $180^\circ$ in steps of $5^\circ$, with $\omega_2 = 90^\circ$ (this corresponds to the stable Kozai libration regime) and $\Delta \Omega = 0^\circ$ (to avoid close encounters). A smaller time step of $\unit[0.35]{d}$ was chosen, which puts the inner boundary (for at least 10 time steps per orbit) at $a_\mathrm{c,1} = \unit[0.055]{AU}$ and therefore allows the possibility of highly eccentric orbits of the test particles, as are expected to occur in Kozai oscillations. The outer boundary $a_\mathrm{c,2}$ again is equal to the semi-major axis of the stellar companion. This time we did not test different binary eccentricities, but only used the best-fit value of 0.93 in our simulations.

The most stable solutions were found for mutual inclinations $135^\circ < \Delta i < 140^\circ$, but even for those configurations the instability timescale is on the order of 10 binary periods ($\cong \unit[500]{yrs}$), after which the test particles are ejected from the system; only a small percentage of systems survive longer than 100 binary periods, but fully stable regions do not appear in the parameter space. We also did not find any signs of Kozai oscillating orbits; if they exist, they must be very short-lived and are unlikely to provide long-term stable solutions.

\section{Possible alternative explanations for the short-period RV variations}\label{SecAlternativeExplanations}

\subsection{Hierarchical triple}\label{SecHierarchicalTriple}

\begin{figure}
\centering
\includegraphics[width=\hsize]{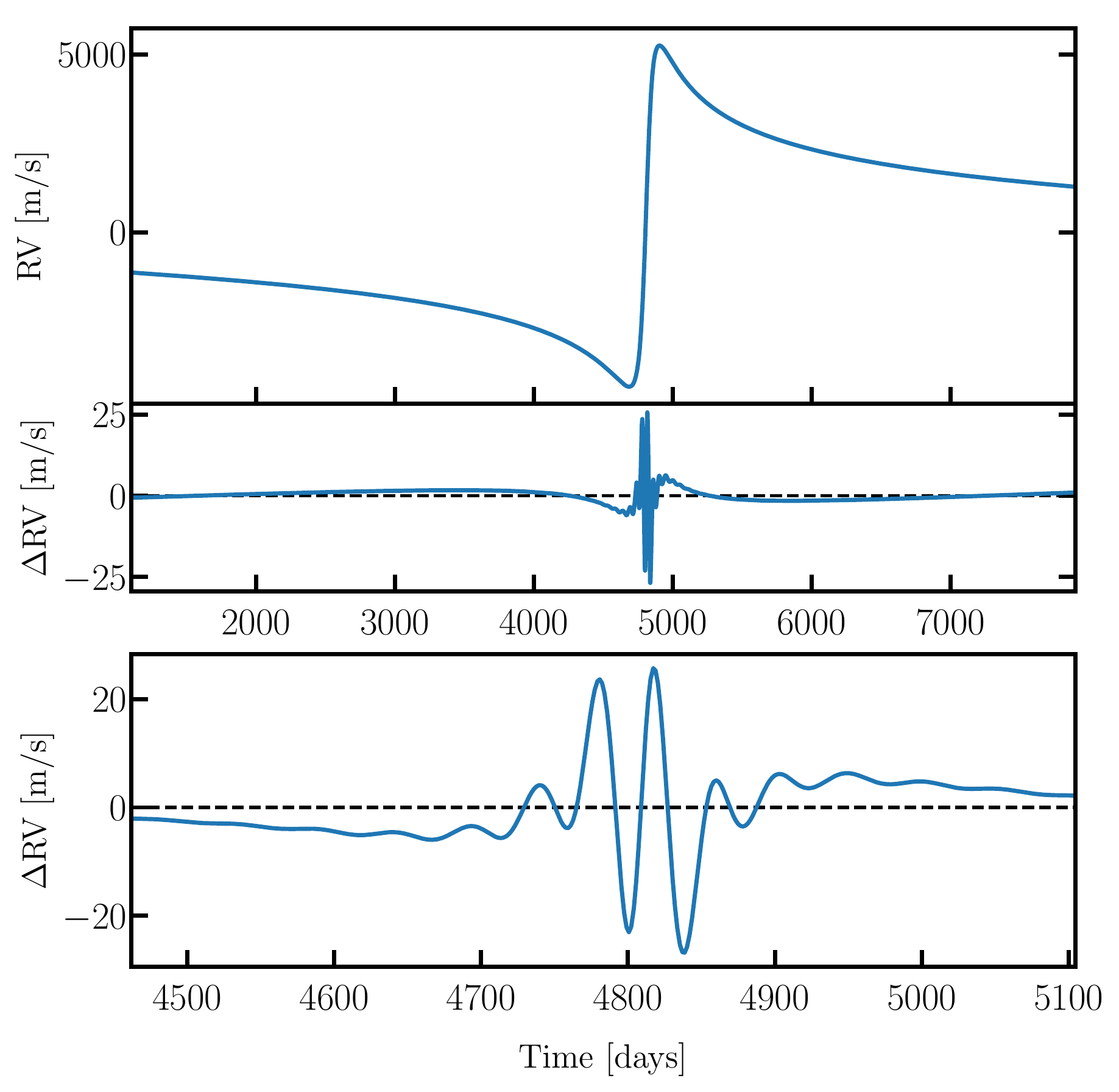}
\caption{\textit{Top:} Simulated RVs of the main component of a hierarchical triple, with the orbit of the companion pair around the primary chosen according to our solution for the close stellar companion to \textepsilon\,Cyg\,A. \textit{Middle:} Residuals of the simulated RVs after subtracting the high-amplitude, long-period RV signal modeled by a Keplerian orbit. \textit{Bottom:} Zoom in on the time around periastron passage of the eccentric long-period orbit.}\label{ImageHierarchicalTriple}
\end{figure}

In a series of publications \citep{Morais2008, Morais2011, Morais2012} it was shown that RV variations of a star can be caused by two bodies orbiting each other while orbiting that star, as is the case in hierarchical triples. In \citet{Morais2012} the authors develop a secular theory for such systems and derive an expression for the precession rate of the orbit of the companion pair around the main component. Furthermore they suggest this effect as an alternative explanation to the planet hypothesis for the $\nu$\,Octantis system \citep{Ramm2009,Ramm2016}, which is known to be a close single-line spectroscopic binary where the two stellar components orbit each other with a period of $\unit[1050]{d}$. In addition to the long-period RV signal caused by the stellar companion, \citet{Ramm2009} observed short-period RV variations with a period of $\unit[417]{d}$, which they interpreted as a possible planet with a minimum mass of $\unit[2.5]{M_{jup}}$. 
Stable orbital configurations have been published by \citet{Ramm2016}, but previously the stability of the proposed planet seemed questionable.
\citet{Morais2012} argue that the short-period RV variations could also be caused by a very close pair orbiting the main component, meaning that the stellar companion $\nu$\,Octantis\,B would itself be made up of two bodies. 
They also show that the precession rate of the long-period orbit as measured from the RVs is in agreement with their hypothesis.

Here we aim to test the hierarchical triple configuration for the \textepsilon\,Cyg system. Unfortunately, our RV data do not even cover one full phase of the long-period orbit, which makes any measurement of the precession rate very unreliable. Therefore we chose a different approach: Using a modified version of the Wisdom-Holman $N$-body integrator in the \textit{SWIFT} package we modeled hierarchical systems and tried to reproduce the RV variations observed in our data. As the integrator was originally created to model the Solar System, it only allows to set up orbits of bodies around the central star of the modeled system, not of two outer bodies orbiting each other whilst orbiting together around the central star. By changing from barycentric to Jacobi coordinates however, we are able to choose one of the two components of the companion pair (denoted as $m_1$) as central object. In this transformed setup, the other component of the companion pair ($m_2$) as well as the main component ($m_0$) are then orbiting $m_1$.

We used a brute-force technique to try to find systems that produce RVs similar to what we observe. The long-period orbit of the companion binary is effectively given by our solution in Sect.~\ref{SecOrbitStellarCompanion}; as described above, these (Jacobian) orbital elements were now ascribed to the main component $m_0$. We then replaced the single companion by a pair of two bodies $m_1$ and $m_2$; $m_1$ was chosen as the central star in the system and we systematically varied the Keplerian elements of the orbit of $m_2$ about $m_1$ as well as the masses of the two bodies. After setting up the system, we let it run for 200 years with a time step of $0.5$ days and recorded the RVs of the main component.

In these simulated RVs the long-period orbit of the $m_1$-$m_2$ pair around the main component $m_0$ was in general clearly visible and closely resembled our data (see top plot of Fig.~\ref{ImageHierarchicalTriple}). We then fitted a Keplerian model to that RV variation and subtracted it from the RVs. 
By computing a GLS periodogram of the residuals, we searched for short-period variations. While in most simulated systems the binarity of the companion was too small to produce any considerable signature, or one of the components was lost within a very short time, some simulated configurations did show peaks at the periods in question. 
These were all systems with a comparably large separation $a_1$ between the two bodies $m_1$ and $m_2$, and all of them showed signs of long-term instability.
However, the short-period variations were always limited to the time around periastron passage of the $m_1$-$m_2$ binary around the main component. 

Figure~\ref{ImageHierarchicalTriple} shows a characteristic example, for a system with $a_1 = \unit[0.3]{AU}$, $m_1 = \unit[0.16]{M_\sun}$ and $m_2 = \unit[0.12]{M_\sun}$: After subtracting the long-period orbit, the residuals show variations with an amplitude of $\unit[30]{m\, s^{-1}}$ near periastron passage (middle plot), with a quasi-periodic nature (lower plot). During that time, all three bodies are very close to each other (the periastron distance of the $m_1$-$m_2$ system to $m_0$ is approximately $\unit[1.2]{AU}$), so the varying quadrupole (and higher) moments of the companion pair induce a large motion on the main component. But this signal quickly decays to amplitudes below $\unit[1]{m\, s^{-1}}$ further away from the time of periastron passage. In contrast, we clearly observe RV variations with amplitudes around $\unit[25]{m\, s^{-1}}$ in the McMillan and Lick data of \textepsilon\,Cyg far from periastron passage.

We conclude that while hierarchical triples may for some systems be a valid alternative explanation for the observed RV signals, this is not a viable solution for the observed RVs of \textepsilon\,Cyg. In contrast to the case discussed in \citet{Morais2012}, the long-period orbit of \textepsilon\,Cyg is far too eccentric to be compatible with large-amplitude short-period RV variations over the whole orbit induced by a companion pair.

\subsection{Stellar spots}\label{SecSpots}

\begin{figure}
\centering
\includegraphics[width=\hsize]{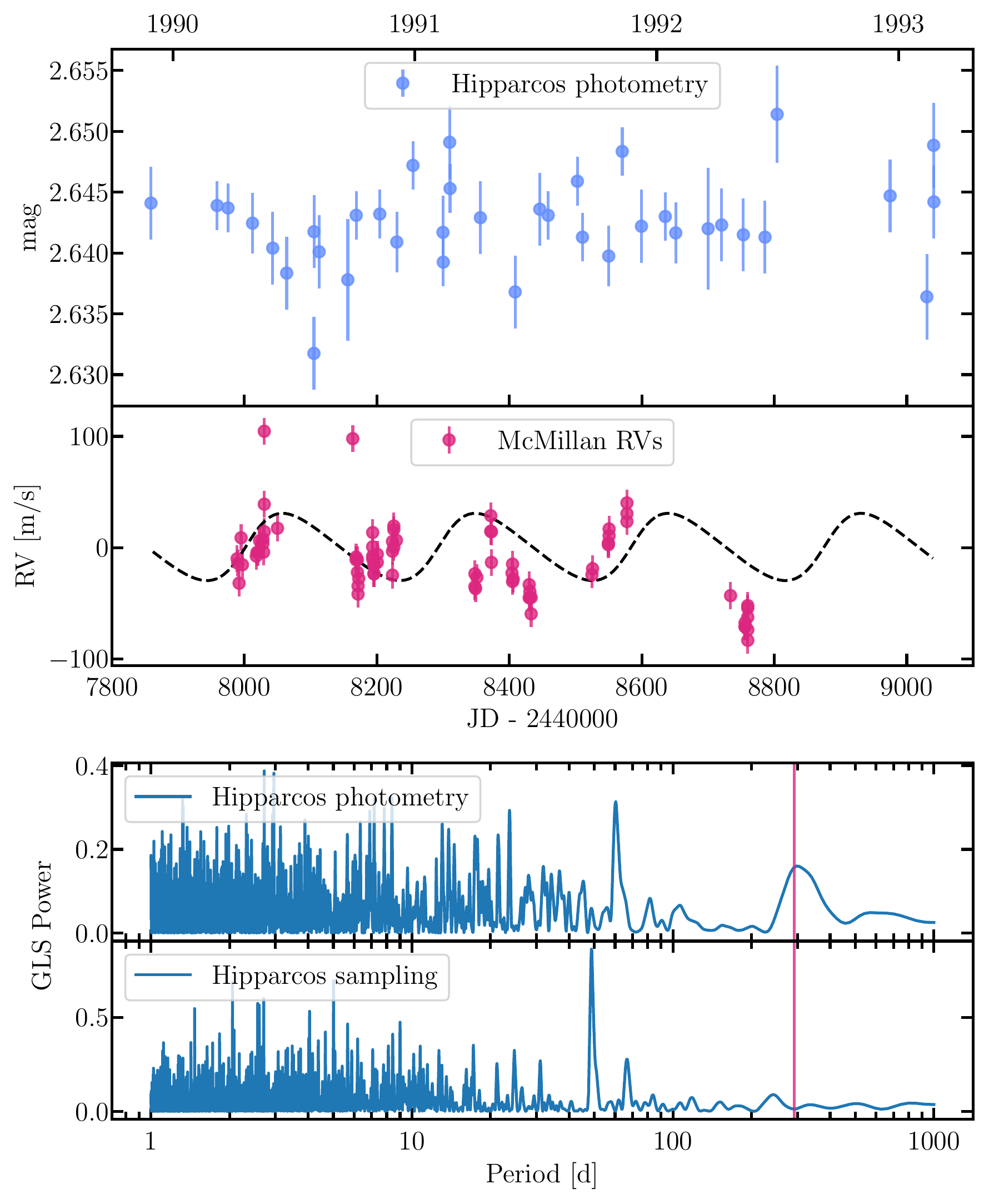}
\caption{\textit{Top:} Hipparcos photometry for \textepsilon\,Cyg, taken around the same time as the last RVs of the McMillan data set. The displayed photometry data are median values for all cases of multiple measurements within one day; the RVs are the residuals of the McMillan measurements after subtracting the long-period signal induced by the binary companion (see Sect.~\ref{SecOrbitStellarCompanion}), and the dashed line denotes the best-fit model. \textit{Bottom:} GLS periodogram of the Hipparcos photometry data and of the window function of the observations. The best-fit Keplerian orbital period of the signal of $\unit[291]{d}$ is marked with a red line.}\label{ImageHipparcosPhotometry}
\end{figure}

\begin{figure}
\centering
\includegraphics[width=\hsize]{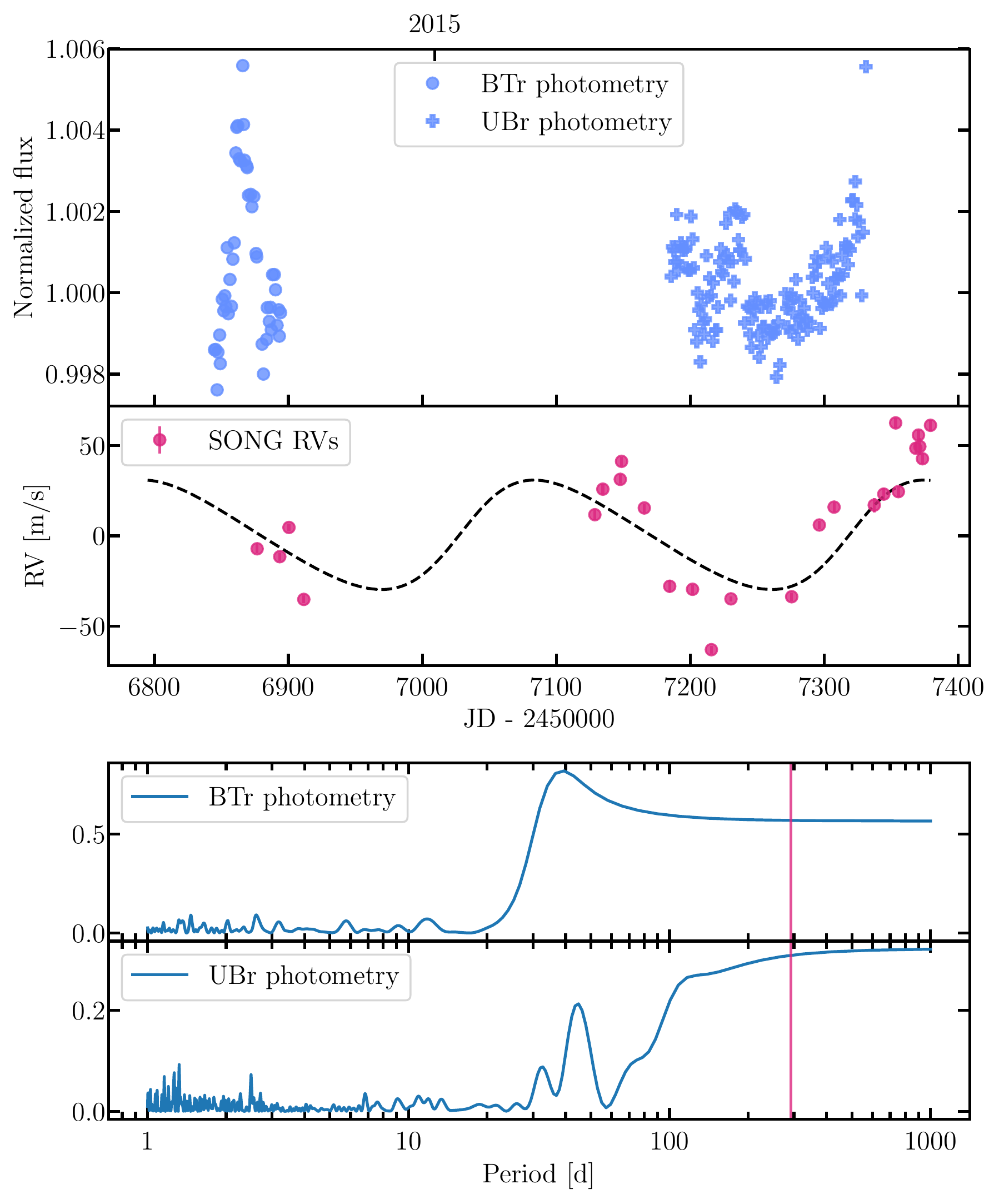}
\caption{\textit{Top}: BRITE-Toronto (BTr) and UniBRITE (UBr) photometry for \textepsilon\,Cyg, taken contemporaneously with the first RVs of the SONG data set. The displayed photometry data are median values for all cases of multiple measurements within one day, and both data sets have been normalized by their respective median values; the RVs are the residuals of the SONG measurements after subtracting the long-period signal induced by the binary companion (see Sect.~\ref{SecOrbitStellarCompanion}), and the dashed line denotes the best-fit model. \textit{Bottom}: GLS periodogram of the Brite photometry data sets. The best-fit Keplerian orbital period of the signal of $\unit[291]{d}$ is marked with a red line.}\label{ImageBritePhotometry}
\end{figure}

Many stars exhibit stellar activity, that is, stellar spots, flares etc., which can mimic a planet in RV observations by blocking or enhancing light from parts of the surface of the stellar photosphere. This effect can become especially problematic when searching for planets around M dwarfs, which are often quite active and therefore can show activity-induced RV signals of several $\unit{m\, s^{-1}}$ \citep[see e.g.,][]{Reiners2010, Barnes2011, Tal-Or2018}. By checking time-series of activity indicators, such as photometry, bisector spans of the absorption lines, or the depths of the H$\alpha$ and Ca\,II lines of these stars, and comparing them to the RV time series one can distinguish between activity-induced and planetary signals. In addition, the so-called chromaticity, that is, wavelength-dependence, of RV signatures caused by stellar activity can serve as a tracer \citep[][]{Reiners2010, Zechmeister2018}.

Evolved stars in contrast are known to be less affected by large-scale activity. Our sample stars were chosen specifically because they were comparably quiet stars in the Hipparcos photometry measurements. The top panel of Fig.~\ref{ImageHipparcosPhotometry} displays the Hipparcos data of \textepsilon~Cyg, which have been averaged whenever there was more than one data point within a day. The time series largely overlaps with the second half of the McMillan RVs, which are displayed in the plot underneath. Apart from some outliers, most of the Hipparcos measurements lie very close around the mean of $\unit[2.643]{mag}$ with an rms scatter of $\unit[3.7]{mmag}$. This translates to a relative change in flux of about $0.34 \%$.

Nevertheless, when taking a periodogram of the Hipparcos photometry, there is a peak around a period of $\unit[300]{d}$ (see bottom panel of Fig.~\ref{ImageHipparcosPhotometry}), which matches the observed period of the RV signal (indicated by the red line) quite well. This might hint at a signal induced by spots in the photometry data and would mean that the rotation period of the star would be approximately $\unit[300]{d}$ or an integer multiple of that. To our knowledge, there are two modern measurements of the projected rotational speed of \textepsilon\,Cyg: \citet{Massarotti2008} determined it to be $v_{\mathrm{rot}} \sin i_\star = \unit[1.2]{km\, s^{-1}}$, while the analysis of \citet{Gray2015} resulted in $v_{\mathrm{rot}} \sin i_\star = \unit[1.0 \pm 0.2]{km\, s^{-1}}$ (where $i_\star$ denotes the inclination of the stellar spin axis). Assuming $i_\star = 90^\circ$ these values correspond to rotation periods of $461$ and $\unit[554 \pm 111]{d}$, so in-between 1 and 2 times the period that is visible in our data. Taking the uncertainty in the measurement of \citet{Gray2015} into account as well as the fact that for smaller values of the unknown inclination $i_\star$ the rotation periods become shorter, it seems possible that the peak in the periodogram of the photometry actually traces the rotation of the star.

To further investigate this we searched the literature for additional photometric time series, but only found two more data sets: The BRITE nano-satellites BRITE-Toronto (BTr) and UniBRITE (UBr) \citep{Weiss2014} observed \textepsilon\,Cyg in the years 2014 and 2015, which coincides with the beginning of our SONG measurements (see Fig.~\ref{ImageBritePhotometry}, top). Unfortunately, the two data sets only cover 50 and $\unit[146]{d}$, respectively, which is much shorter than one full phase of the short-period signal and therefore does not help to shed more light on the question whether that signal has a direct counterpart in photometry (see Fig.~\ref{ImageBritePhotometry}, top). The rms scatter of the data sets BTr and UBr are 2.0 and $\unit[1.1]{mmag}$, respectively, which is even lower than in the Hipparcos data. Interestingly however the periodograms of the measurements reveal some variability on even shorter periods, with a strong peak around $\unit[40]{d}$ in both data sets (see Fig.~\ref{ImageBritePhotometry}, bottom). The Hipparcos periodogram seems to have a rough analog with a peak around $\unit[60]{d}$, although in that case it might also be caused by the sampling frequency (see Fig.~\ref{ImageHipparcosPhotometry}, bottom).

Generally, it would be surprising if the observed short-period RV variations of \textepsilon\,Cyg were caused by spots, as it is an evolved star with a radius of about $\unit[11]{R_\sun}$. In order to cause RV variations on the order of the observed semi-amplitude of $\unit[20 \sim 30]{m\, s^{-1}}$, stellar spots would have to be extremely large, which in turn should become clearly visible in photometry as well. To study the effect qualitatively, we used the online tool of the star spot simulator \textit{SOAP 2.0} \citep{Dumusque2014} to find a combination of spots that could possibly mimic the observed RV signal. Setting the rotation period of the simulated \textepsilon\,Cyg\,A to the observed modulation of $\unit[291]{d}$, we found that a single spot would need to have a radius of $\unit[0.2]{R_\star}$ to induce a RV semi-amplitude of about $\unit[25]{m\, s^{-1}}$, and the resultant flux change would be around $2.7 \%$, much more than the scatter of $0.34 \%$ in the Hipparcos photometry. Using more than one single spot to create the same RV amplitude generally allows to decrease the individual spot sizes a little, but the photometric variability remains the same or even increases. Four spots with radii of $\unit[0.12]{R_\star}$ each at very close longitudes and latitudes, for example, keep the RV semi-amplitude roughly constant but lead to a flux variability of $3.5 \%$.
Similarly, placing spots on opposite sides of the star, or assuming an inclination $i < 90^\circ$, only increases the predicted photometric variability.

We conclude that even though there are hints of a signature of the rotation of the star in the Hipparcos periodogram, the modeling suggests that spots are an unlikely cause of the observed RV variations.

\subsection{Oscillatory convective modes}\label{SecOscillatoryConvectiveModes}

In many evolved stars with very high luminosities, photometric oscillations are known with typical periods between about 400 to $\unit[1500]{d}$, but the causes of these oscillations are not fully understood \citep{Wood1999, Hinkle2002, Wood2004}. Those stars are classified as long secondary period (LSP) variables because their primary oscillation periods are much shorter, with the period ratios between the secondary and primary oscillations ranging from about 5 to 13 \citep{Wood1999}. While the oscillations in most LSP stars have only been detected photometrically, \citet{Hinkle2002} and \citet{Wood2004} also obtained RVs for some of these stars and found the long variations present in most cases, with amplitudes of a few $\unit[]{km\, s^{-1}}$ and periods consistent with the photometric variations. \citet{Saio2015} examine the observed periodic variations of the LSP stars and suggest so-called oscillatory convective modes, that is, nonadiabatic g$^-$ modes in the deep convective envelopes of the stars, as a possible explanation.

However, all of the known LSP stars have luminosities of $L_\star > \unit[300]{L_\sun}$ \citep{Saio2015}, whereas \textepsilon\,Cyg has a much lower luminosity of $L_\star = \unit[57]{L_\sun}$. It is therefore not clear whether the mechanism described by \citet{Saio2015} might be responsible for the observed RV variations in this case, but it is certainly possible that LSP oscillations also exist in stars with lower luminosities, and that the distribution of known LSP stars is constrained by observational biases: Most of them have been found and characterized in the \textit{OGLE} survey \citep[see e.g.,][]{Udalski1997,Soszynski2009}, which is aimed at discovering microlensing events and focuses on very distant stars in fields toward the Galactic Center and the Large Magellanic Cloud; stars similar to \textepsilon\,Cyg therefore are probably just not bright enough to be observed with sufficient S/N in order to detect the photometric variations.

Two other examples of evolved stars whose RV variations might well be caused by this mechanism are \textgamma\,Draconis \citep{Hatzes2018} and Aldebaran \citep{Reichert2019}; both show signals with periods around 700 and $\unit[600]{d}$, respectively, which could easily be confused with Keplerian signals if one just examined parts of the data. Similar to \textepsilon\,Cyg, in both cases only the very long time span of the collected RV measurements allowed the authors to discover the varying nature of the signal and therefore reject the planet hypothesis. As the luminosities of \textgamma\,Draconis and Aldebaran are $L_\mathrm{\gamma Dra} = \unit[510 \pm 51]{L_\sun}$ \citep{Hatzes2018} and $L_\mathrm{Ald} = \unit[402 \substack{+11\\-10}]{L_\sun}$ \citep{Reichert2019}, they fall comfortably inside the distribution of known LSP stars and therefore oscillatory convective modes seem to be a plausible explanation for the observed RV signals. With only these two examples of similar (non-)detection histories, it could be that \textepsilon\,Cyg is another example of that category, even though its stellar parameters might suggest otherwise.

\subsection{Potential stellar oscillations through the heartbeat phenomenon}\label{SecStellarOscillations}

A different mechanism that is known to produce oscillations in some eccentric binary systems is the so-called heartbeat phenomenon, where a stellar companion passes close by the primary component and excites tides within the latter. With the \textit{Kepler} mission alone, more than 150 of these systems have been observed \citep{Kirk2016}. They are characterized by a sudden increase in brightness of the primary star during periastron passage of the companion, due to the deformation of the star caused by the increased gravitational force, which is called the equilibrium tide. Furthermore, sometimes one or even several periodic luminosity-signals can be observed over the whole orbit, the so-called dynamical tides. These oscillate in characteristic stellar oscillation modes, which are excited through the regular close passage of the stellar companion \citep[for a full theoretical description see e.g.,][]{Fuller2017}.

\begin{figure}
\centering
\includegraphics[width=\hsize]{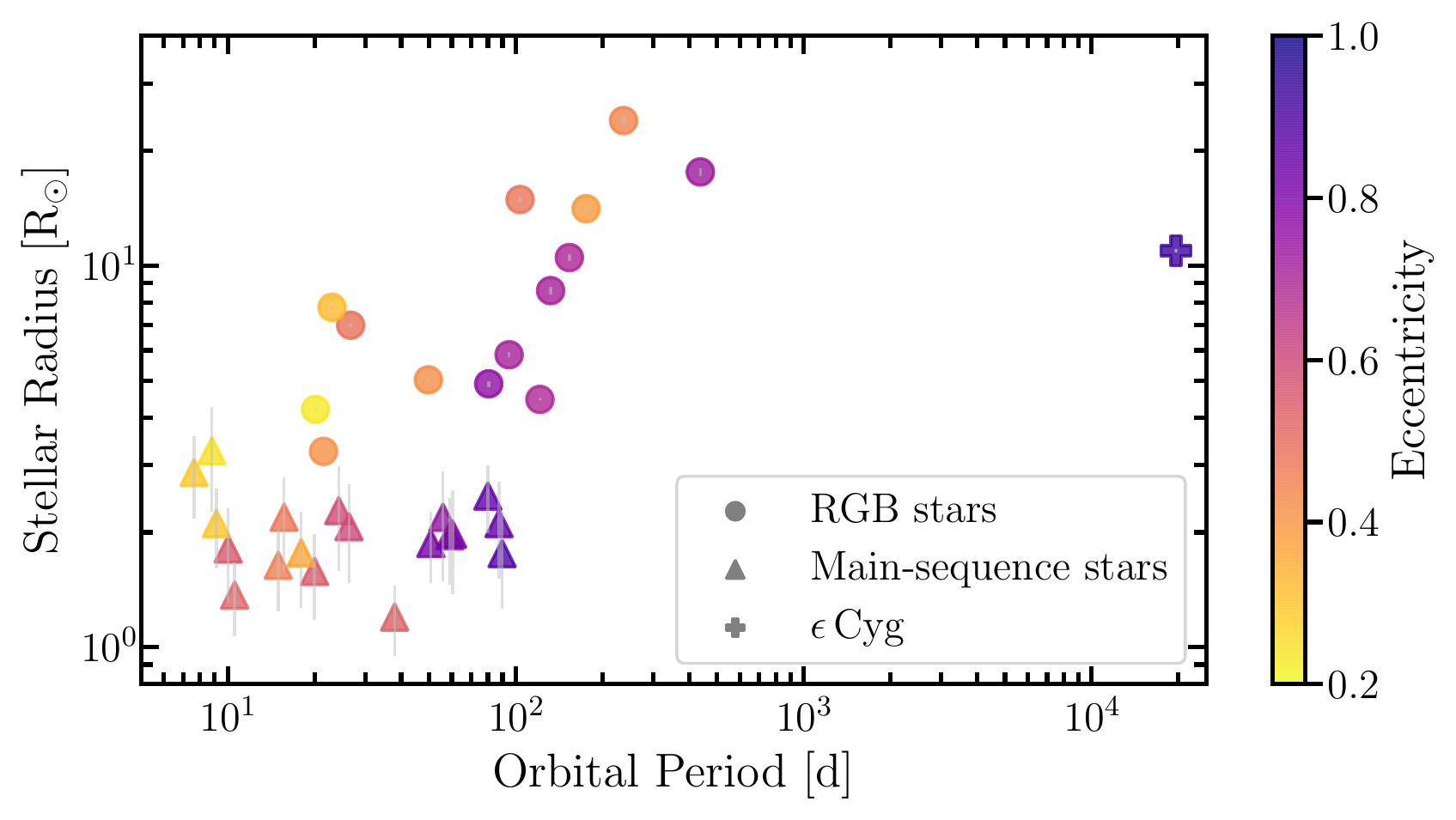}
\caption{Primary stellar radii over binary orbital periods for a number of heartbeat systems, with the orbital eccentricity color-coded. Triangles mark the main-sequence stars from \citet{Shporer2016}, circles the RGB stars from \citet{Beck2014}. For comparison \textepsilon\,Cyg has been added, denoted by a cross. The plot has been created in analogy to \citet{Beck2014}.}\label{ImageHeartbeatstarsBeck}
\end{figure}

All of the known binary systems with heartbeat variations are however much more compact than \textepsilon\,Cyg, with orbital periods typically shorter than 1\,yr \citep{Kirk2016}. Still, in an attempt to investigate this possibility further, we revisit two publications on heartbeat systems that have been monitored in RVs to better constrain their orbits: \citet{Shporer2016} analyze a sample of 19 main-sequence stars in heartbeat binaries, with orbital periods between $8 - \unit[90]{d}$, while \citet{Beck2014} examine 18 RGB heartbeat stars whose stellar companions orbit them with periods between $20 - \unit[438]{d}$. In analogy to Fig.~18 from \citet{Beck2014}, Fig.~\ref{ImageHeartbeatstarsBeck} displays important orbital and stellar parameters of all systems from both samples. The plot shows a clear positive correlation between primary stellar radius and binary orbital period for the RGB stars, while the main-sequence stars seem to be scattered more or less uniformly at the small-radius, short-period end of the distribution. Generally, for a given primary stellar radius heartbeats may occur in systems with longer orbital periods $P$ if the eccentricity $e$ is also larger, as the periastron distance scales as $q \propto P^{2/3} (1-e)$.
\textepsilon\,Cyg however falls far away from all other systems at a much longer orbital period; even though its eccentricity is also larger than any other in the combined sample and its stellar radius rather large, it is questionable from this figure whether \textepsilon\,Cyg could be a heartbeat system.

\begin{figure}
\centering
\includegraphics[width=\hsize]{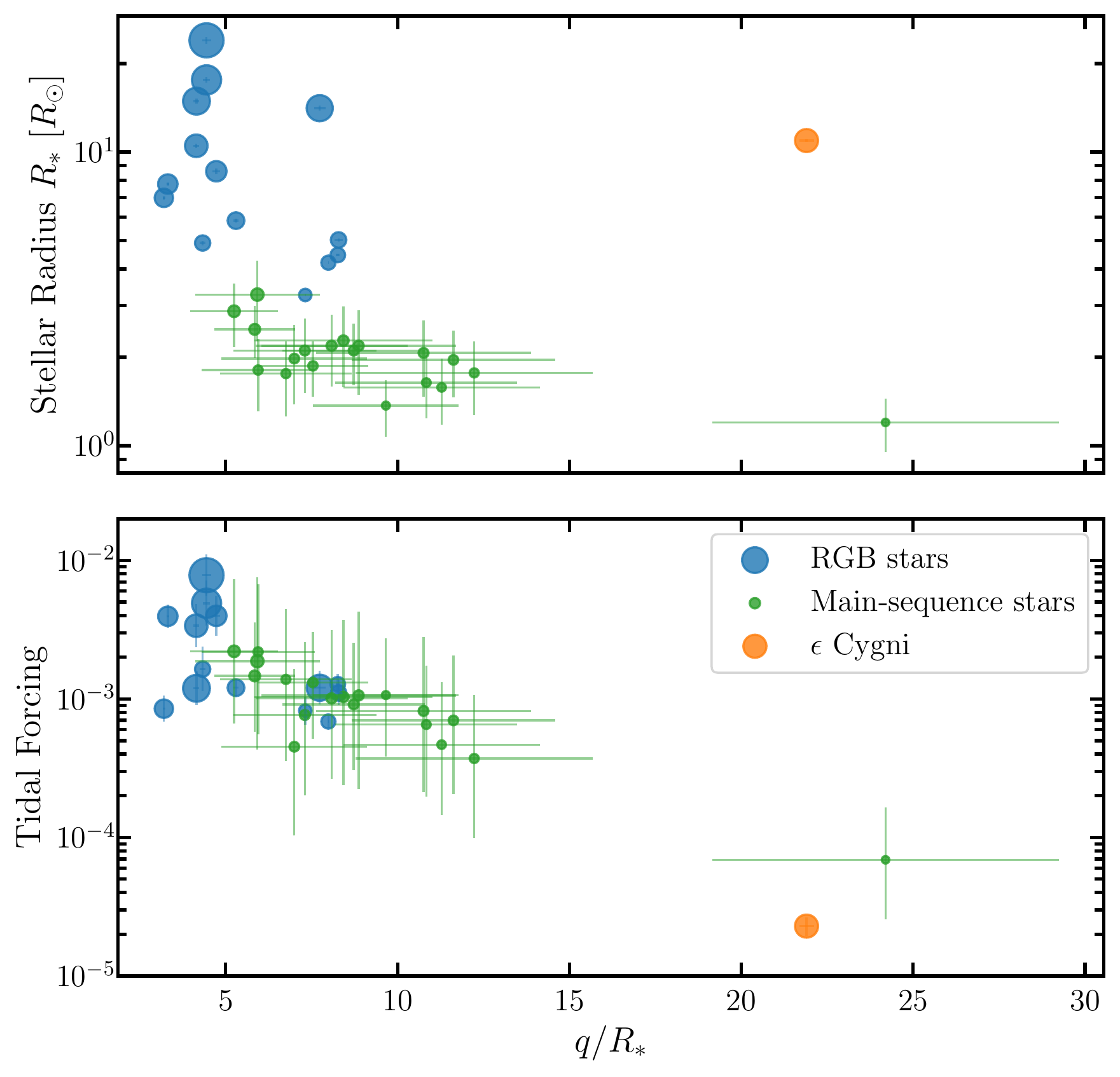}
\caption{Primary stellar radii (top) and tidal forcing ratio (bottom) over periastron distances in units of stellar radii ($q / R_*$) for heartbeat systems, consisting of the main-sequence star sample from \citet{Shporer2016} (green) and the RGB stars from \citet{Beck2014} (blue). \textepsilon\,Cyg has been added for comparison (orange). Symbol sizes are proportional to the stellar radii. The tidal forcing ratio is calculated as in Eqn.~\ref{Tidalforcing}, and the large errors of the main-sequence stars stem from the high uncertainties on the radii $R_1$ and the masses $M_1$ of the primary components. The plot has been created in analogy to \citet{Shporer2016}.}\label{ImageHeartbeatstarsShporer}
\end{figure}

\citet{Shporer2016} further investigate the ratio of the tidal forcing due to the stellar companion to the surface gravity of the primary, using the relation
\begin{equation}\label{Tidalforcing}
\frac{F_\mathrm{tide}}{F_\mathrm{gravity}} =  \left(\frac{G R_1 M_2}{q^3}\right) \left(\frac{G M_1}{R_1^2}\right)^{-1} = \left(\frac{R_1}{q}\right)^3 \frac{M_2}{M_1} \, ,
\end{equation}
where $R_{1,2}$ and $M_{1,2}$ are the radius and mass of the primary and secondary component, respectively, $G$ is the gravitational constant and $q$ is the periastron distance of the two stars. To compute the masses of the secondary components, the authors used the mass functions from their RV analysis and assumed $\sin^3 i = 0.6495$; this value corresponds to the median of the $\sin^3 i$ distribution for the unknown inclinations and it was applied instead of the mean due to the very asymmetric shape of the distribution \citep[see][for more information]{Shporer2016}. We redid these calculations for both samples of \citet{Shporer2016} and \citet{Beck2014}, using the stellar and orbital parameters from their works, as well as for the \textepsilon\,Cyg system (scaling its companion mass $M_2$ by the same inclination as used for the other systems). In Fig.~\ref{ImageHeartbeatstarsShporer} we plot the primary stellar radius (top) and tidal forcing ratio as calculated in Eqn.~\ref{Tidalforcing} (bottom) over the periastron distance in units of the primary stellar radius ($q / R_*$) for each system; the stellar radii are also indicated by the sizes of the symbols in the plots.

Plotted against $q / R_*$, the stars of the two samples now fall much closer to each other, with most of them having periastron distances between 4 and $\unit[12]{R_*}$. The RGB heartbeat stars populate the area at small values of $q / R_*$, whereas the main-sequence stars extend the distribution to larger values of $q / R_*$. One main-sequence heartbeat system even has a $q \approx \unit[24]{R_*}$ (this one being KIC\,10334122), which is comparable to the periastron distance of \textepsilon\,Cyg, $\unit[22]{R_*}$. Due to its larger radius however, \textepsilon\,Cyg falls above the distribution of the samples of RGB and main-sequence stars.

Finally, as is to be expected from Eqn.~\ref{Tidalforcing}, the ratio of the tidal forcing to the surface gravity of the primary shows a clear correlation with $q / R_*$. Most of the main-sequence and RGB stars fall between values of $4 \cdot 10^{-3}$ (when $q / R_*$ is small) and $4 \cdot 10^{-4}$ (when $q / R_*$ is larger), and the rms scatter of the two samples combined is $1.55 \cdot 10^{-3}$; the main-sequence system at $q \approx \unit[24]{R_*}$ has a tidal forcing ratio around $7 \cdot 10^{-5}$. 
Calculating the same quantity for \textepsilon\,Cyg gives a value of $2.3 \cdot 10^{-5}$, which is not far from the main-sequence outlier. Given that \textepsilon\,Cyg lies so close to a known heartbeat system in these quantities makes it appear plausible that its primary star \textepsilon\,Cyg\,A might also undergo tidally induced stellar oscillations. 
Additionally, it is important to keep in mind that the tidal forcing ratio scales with the companion mass $M_2$, and that \textepsilon\,Cyg might therefore be even closer to the other stars in the case of a low inclination of the system. For example, for the lower limit on the inclination of $i = 14^\circ$, the tidal forcing would be $1 \cdot 10^{-4}$, which is greater than that of the main-sequence outlier and nearly within the rms scatter from the mean of the RGB and main-sequence distribution.

We also emphasize the fact that the variability of \textepsilon\,Cyg has been observed in RVs, whereas the systems from \citet{Beck2014} and \citet{Shporer2016} have all been discovered through their photometric variations in Kepler observations. Even though both these studies also monitored their systems in RVs to constrain their orbits, they do not mention any detection of RV variations corresponding to the heartbeat oscillations. This could have several reasons: They might not have picked up the signals in RVs due to sparse sampling or large errors of their measurements. Also some of their heartbeat systems barely seem to show any noticeable dynamical tides and mostly just equilibrium tides during periastron passage of the stellar companion, so there might not even be an RV signal to detect during most parts of the orbit.
For \textepsilon\,Cyg in contrast we have a large number of RV measurements over most parts of the binary orbit, which clearly record a short-period signal that changes in period and semi-amplitude especially around periastron passage of the stellar companion. Given the differing detection method as compared to the other heartbeat stars, it is possible that \textepsilon\,Cyg experiences a similar phenomenon even though it does not follow all correlations exactly.

\section{Summary \& conclusions}\label{SecSummaryConclusion}

Short-period RV variability of the spectroscopic binary \textepsilon\,Cyg had already been noticed in past publications \citep{McMillan92, Gray2015} and had been attributed to intrinsic stellar variations rather than an extrinsic source. We observe a similar RV signal in our measurements of the star at Lick and with SONG and further investigated the possibility of a planetary origin.

In a first step we determined the long-period binary orbit with very high precision by fitting a Keplerian model to the three RV data sets of \citet{McMillan92}, Lick and SONG. The results indicate the possibility of directly imaging the secondary component in the years 2021/2022. A direct observation would  constrain the orbital inclination and the mass of the spectroscopic companion, possibly revealing whether it is a white dwarf or a main-sequence star, which could give insight into the past evolution of the binary system.

Next we computed GLS periodograms of the residuals after subtracting the Keplerian model from the RV data; they show clear signs of additional periodic signals at periods just below $\unit[300]{d}$ for all three data sets. We fitted the combined RV data with a double-Keplerian model to explore the possibility of an S-type planetary companion as cause for the short-period RV variations. The best fit leaves a considerable number of systematic outliers that look like a phase shift of the signal at some point in time or a change of orbital period. By modeling each data set, and smaller sections of the data sets individually, we revealed that the planetary period gradually increases over time, whereas the RV semi-amplitude mostly stays constant until it increases by a factor of 2 at periastron passage of the stellar companion. This indicates that a single circumprimary planetary companion is an unlikely explanation for the observed RV signal, but the residuals of the double-Keplerian model could point toward a scenario with a second low-mass, co-orbital companion in the system. Using a fully dynamical model for the planetary companion, treating it as a test particle in the stellar binary system, does not help to solve the issue, as it cannot reproduce the systematic changes of the Keplerian elements, neither for the prograde nor for the retrograde case. Additionally, our stability analysis shows highly unstable behavior in large regions around the best-fit orbital configurations, thus strengthening the arguments against a planet.

Following \citet{Morais2008, Morais2011, Morais2012}, we explored whether the RV variations could be due to an additional component not orbiting the primary, but the secondary star, making the system a hierarchical triple. Our models clearly show that in our case such a setup cannot reproduce the data. We therefore investigated possible intrinsic stellar origins of the signal, such as stellar spots. Data from Hipparcos suggest that \textepsilon\,Cyg\,A is a photometrically relatively quiet star, even though a periodogram of the measurements reveals a possible weak signal around $\unit[300]{d}$, which is close to the period observed in RVs and could correspond to the rotation period of the star. If the RV variations were due to stellar spots, we would however expect the photometric variation to be much larger than the observed scatter.

Two other examples of ambiguous RV variations of giant stars, \textgamma\,Draconis \citep{Hatzes2018} and Aldebaran \citep{Reichert2019}, might belong to the class of LSP variables described by, for example, \citet{Hinkle2002}, \citet{Wood1999,Wood2004}, and \citet{Saio2015}. As \textepsilon\,Cyg\,A is much smaller and less luminous than the known LSP stars, this mechanism can offer a satisfactory explanation in this case only if the distribution of LSP stars known today is incomplete due to observational biases.

A more likely, even though also not completely convincing solution is the possibility of \textepsilon\,Cyg being a heartbeat system, with the close stellar companion inducing tides in the primary component during each periastron passage that excite long-standing oscillations with the observed period. However, there are some distinct differences between the known heartbeat systems and \textepsilon\,Cyg, with the periastron distance of the latter being much larger. Also the question remains whether there are oscillation modes at periods around $\unit[300]{d}$ in HB stars that could easily be excited through this mechanism.

Generally, we showed that none of the phenomena discussed above offer a fully satisfactory explanation for the RV variability by themselves at the moment. Nevertheless, it is possible that a combination of some of the mechanisms could explain the observed RVs while also eliminating some of the problems encountered in the analysis. For example, the oscillatory convective modes might be present in \textepsilon\,Cyg\,A and, although not strong enough to cause recognizable RV variations by themselves in this type of star, could be excited regularly by the tidal interaction with the stellar companion. On the one hand, this would explain why \textepsilon\,Cyg\,A is so far away from the distribution of known LSP stars and still shows oscillations; on the other hand, the fact that these oscillation modes are present and probably easily excited offers a solution to the problem of the comparably large separation of the stellar binary when compared to other heartbeat systems. Similarly, \citet{Saio2018} show the presence of tidally excited Rossby waves (so-called r modes) in some heartbeat systems, which typically appear at periods slightly longer than the rotation period. If the observed RV signal of \textepsilon\,Cyg was caused by this mechanism, the star's rotation period therefore would have to be shorter than $\sim\unit[280]{d}$, and the inclination of its spin axis $i_\star$ smaller than $\sim 30^\circ$ (see Sec.~\ref{SecSpots}, given that the $v_\mathrm{rot}$ measurement by \citet{Gray2015} is correct). As the work by \citet{Saio2018} is focused on upper MS stars, it obviously remains questionable whether r modes can also be present in HB stars like \textepsilon\,Cyg\,A.

Another possibility is that there actually is a planet present around \textepsilon\,Cyg\,A, and its orbital period corresponds to the rotational period of the star, meaning the system is tidally locked. The observed RV variations could then be a sum of the Keplerian signal of the star and the rotational modulation, therefore solving the mismatch between the low photometric variations and the large RV amplitude for a rotational signal alone. Additionally, a tidally coupled system might improve the stability of a planetary companion in an orbit around the best-fit parameters derived in Sect.~\ref{SecOrbitBothCompanions}. In order to investigate this, it is necessary to pair an $N$-body code with a state-of-the-art algorithm for the computation of tidal forcing, which is beyond the scope of this work.

For the sake of completeness, we also mention the recent work by \citet{Maciejewski2020}, who show that the slightly eccentric RV signature of WASP-12, which has been explained by a hot Jupiter on an unusual eccentric orbit, could also be explained by a circular orbit of the planet and the signature of tides in the host star. \citet{Arras2012} delivered the theoretical foundation, proving that tides in stars, induced by massive planets on tight orbits, can manifest themselves in RV signals with periods of half the orbital period and an amplitude of a few $\unit[]{m\, s^{-1}}$; due to a phase shift between the planetary and tidal RV signals the combined signature can then mimic an orbit with nonzero eccentricity and longitude of periastron equal to $270^\circ$. This raises the question whether a similar effect could be at work in \textepsilon\,Cyg, where tides raised in the primary \textepsilon\,Cyg\,A during the close periastron passage of the spectroscopic companion might contribute to the recorded RV signature; with the eccentricity of the long-period orbit being constrained mostly by the data around periastron passage, the true orbital eccentricity might then be lower than our models suggest. This in turn could solve the stability issues of the putative planet discussed in Sect.~\ref{SecStabilityAnalysis} if the eccentricity actually would drop below a value of $0.9$. It is questionable though whether the RV contribution of tides would be large enough, and a more thorough analysis is needed to assess this possibility.

Generally however, we must conclude that the cause of the short-period RV variations of \textepsilon\,Cyg remains a mystery for now. This analysis confirms what has been shown in other recent publications about evolved stars: With more years of continued RV measurements we are now detecting more stars with signals that mimic planets but are more likely caused by other phenomena. Only thanks to the long time spans of the observations and through careful analysis are we able to identify these false positives. This should be a warning, as a considerable number of the published planets around giant stars, especially the early discoveries, had been declared on the basis of very sparse data sets, and some of them might have to be revoked once more measurements are available.

\begin{acknowledgements}
Part of this work was supported by the International Max Planck Research School for Astronomy and Cosmic Physics at the University of Heidelberg, IMPRS-HD, Germany.  
We would like to thank the staff at the Lick Observatory for their support during the years of this project. 
We are also thankful to the CAT observers who assisted with this project, including Saskia Hekker, Simon Albrecht, David Bauer, Christoph Bergmann, Stanley Browne, Kelsey Clubb, Dennis K\"ugler, Christian Schwab, Julian St\"urmer, Kirsten Vincke, and Dominika Wylezalek.  
This work also includes observations made with the Hertzsprung SONG telescope operated at the Spanish Observatorio del Teide on the island of Tenerife by the Aarhus and Copenhagen Universities and by the Instituto de Astrof\'isica de Canarias.
Part of the work has been performed with the RV analysis tool Exo-Striker (\url{https://ascl.net/1906.004}), and many of the simulations have been carried out on the machine of the {\it Novel Astronomical Instrumentation based on photonic light Reformatting} (NAIR) group at the Landessternwarte Heidelberg. 
This research has been funded by the Spanish State Research Agency (AEI) Projects No.ESP2017-87676-C5-1-R and No.\ MDM-2017-0737 Unidad de Excelencia ``Mar\'ia de Maeztu''- Centro de Astrobiolog\'ia (INTA-CSIC). 
Funding for the Stellar Astrophysics Centre is provided by The Danish National Research Foundation (Grant agreement no.: DNRF106). 
SA acknowledges the support from the Danish Council for Independent Research through the DFF Sapere Aude Starting Grant No.\ 4181-00487B. 
K.H.W. and M.H.L. are supported in part by Hong Kong RGC grant HKU 17305015.
VA was supported by a research grant (00028173) from VILLUM FONDEN.
The research made use of the SIMBAD database operated at CDS, Strasbourg, France, as well as of the NASA Exoplanet Archive, which is operated by the California Institute of Technology, under contract with the National Aeronautics and Space Administration under the Exoplanet Exploration Program.
We are grateful to Bogdan Ghiorghiu's contribution in the analysis of the RV measurements. 
We thank A.\ Leleu and A.\ Correia for private discussions on the co-orbital demodulation technique. 
We also thank H.\ Kjeldsen for his feedback concerning the jitter analysis, and R.\ Kuschnig for making the BRITE data available to us.
Finally, we thank the anonymous referee for his very valuable review, especially his questions concerning the jitter analysis.
\end{acknowledgements}

\bibpunct{(}{)}{;}{a}{}{,} 
\bibliographystyle{aa} 
\bibliography{references}

\end{document}